\documentclass[aps,twocolumn]{revtex4}
\usepackage [cp1251]{inputenc}
\usepackage{amsmath}
\usepackage{bm}
\usepackage{graphicx}
\usepackage{bm}
\usepackage{epsfig}
\usepackage{epstopdf}
\usepackage{amsmath}
\usepackage{color}

\begin{document}
\title{Theory of single and double electron spin-flip Raman scattering\\ in  semiconductor nanoplatelets}

\author{A.V. Rodina\footnote{anna.rodina@mail.ioffe.ru} and E.L. Ivchenko}

\affiliation{Ioffe Institute, 194021 St.~Petersburg, Russia}

\begin{abstract}
A theory of electron spin-flip Raman scattering (SFRS) is presented that describes the Raman spectral signals shifted by both single and twice the electron Zeeman energy under nearly resonant excitation of the heavy hole excitons in semiconductor nanoplatelets. We analyze the spin structure of photoexcited intermediate states, derive compound matrix elements of the spin-flip scattering and obtain polarization properties of the one- and two-electron SFRS common for all the intermediate states. We show that, in the resonant scattering process under consideration, the complexes ``exciton plus localized resident electrons'' play the role of main intermediate states rather than tightly bound trion states. It is demonstrated that, in addition to the direct photoexcitation (and similar photorecombination) channel, there is another indirect channel contributing to the SFRS process. In the indirect channel,  the photohole forms the exciton state with the resident electron removed from the   localization site while the photoelectron becomes localized on  this site. The theoretical results are compared with recent experimental findings for ensembles of CdSe nanoplatelets.

%71.35.--y (Excitons and related phenomena), 78.55.Et (II-VI semiconductors), 78.30.--j (Infrared and Raman spectra),  78.67.--n (Optical properties of low-dimensional, mesoscopic, and nanoscale materials and structures),
\end{abstract}
%\pacs{71.35.--y, 78.55.Et, 78.30.--j, 78.67.--n}

\date{\today}
  \maketitle

\section{Introduction}
Spin-flip Raman scattering is an electronic process of inelastic light scattering with the initial and final states being the different spin-states of electrons and/or holes. In semiconductors, spin-flip Raman scattering (SFRS) was predicted by Yafet \cite{Yafet} in 1966 and first observed by Slusher, Patel and Fleury \cite{Patel} in InSb. The observed Raman scattering mechanisms in semiconductors and semiconductor nanostructures include spin flip of mobile carriers \cite{Yafet,Patel}$\mbox{}$, resident and photoexcited, or carriers bound to shallow donors and acceptors (via exciton-involved processes) \cite{ToHo1968,ScottReview,SaCa1992}$\mbox{}$ as well as spin flip of excitons mediated by bulk acoustic vibrations \cite{Sirenko1998}$\mbox{}$.

In addition to numerous publications on the single-electron SFRS, there are few publications on the double spin-flip Raman scattering in bulk semiconductors, namely, CdS \cite{Scott1972}$,$ see also Refs.~[\onlinecite{ToHo1968,ScottReview}], and ZnTe \cite{doubleCdTe,OkaCardona}. In Ref.~\cite{OkaCardona}, in addition to single- and double-electron SFRS, there was observed a triple spin-flip scattering process in which three spins of donor electrons are reversed.

Recently, this kind of scattering with a Raman shift twice the single spin-flip energy has been observed in CdSe nanoplatelets \cite{Kudlacik2020}$\mbox{}$, a new type of two-dimensional nanocrystals that emerged a decade ago~\cite{Ithurria2008}$\mbox{}$. In Ref.~[\onlinecite{Kudlacik2020}] we have also proposed a theory to explain the experimental findings, first of all the polarization properties of the SFRS. In this paper we extend a brief theoretical consideration of Ref.~[\onlinecite{Kudlacik2020}] to a full scale presentation of the theory of SFRS mediated by excitons interacting either with one or two resident electrons localized in a nanoplatelet (NPL).

The theory of multiple spin-flip Raman scattering proposed by Economou et al. \cite{Economou1972}~and published in the same issue of the Physical Review Letters as the experiment \cite{Scott1972}~ is based on the exchange interaction of two or more donor-bound electron spins with the electron spin in the photoexcited exciton. Here we extend the theory to consider the single and double spin-flip scattering processes in colloidal nanoplatelets hosting more than one localized resident electron. Moreover, as compared to Ref.~[\onlinecite{Economou1972}] we analyze the compound matrix elements of the spin-flip scattering taking into account the spin and orbital structure of the resonant intermediate states, derive simple polarization properties of the one- (1$e$) and two-electron (2$e$) SFRS common for all intermediate states and discuss reasons for violation of these selection rules.

The rest of the paper is organized as follows. In Sect.~\ref{setup}, we describe the geometrical set-up of the spin-flip Raman scattering process, the Zeeman states of one and two resident electrons in the NPLs as well as the exciton eigenstates. In Sect.~\ref{single}, we analyze the structure of the three-particle intermediate state (\ref{states}), derive the general expression for the single SFRS compound matrix element (\ref{matrix}) and consider two limiting cases corresponding to the trion and ``exciton plus localized resident electron'' intermediate states (\ref{limits}). We discuss in Sect.~\ref{indirect} the origin of different contributions to the single SFRS coming from direct and indirect photoexcitation and recombination channels.  In Sect.~\ref{double}, we derive the compound matrix elements for the double SFRS mediated by the complex ``exciton plus two localized resident electrons'' including the direct (\ref{Doublegeminate}) and indirect (\ref{doubelindirect}) channels as well as by the ``singlet trion plus  localized resident electron'' complex (\ref{doubletrion}). Sect.~\ref{compratio} presents a derivation of simplified expressions for the resident electron, exciton and trion wave functions, allowing us to obtain estimations for the efficiency of different mechanisms of SFRS and compare the theory with  experiment. The discussion of polarization selection rules and their comparison with the experimental observation is given in Sect.~\ref{polar}. Finally, we summarize our findings in Sect.~\ref{conclusion}. The expressions for the absorption and emission matrix elements are presented in the Appendix~\ref{AbsEmiss}. The Appendix~\ref{B} contains integral relations used in the estimations of Section~\ref{compratio}.

 \section{The scattering set-up and Zeeman splitting} \label{setup}
 \begin{figure}[h] 	
 	\centering
 		\includegraphics[width=8 cm]{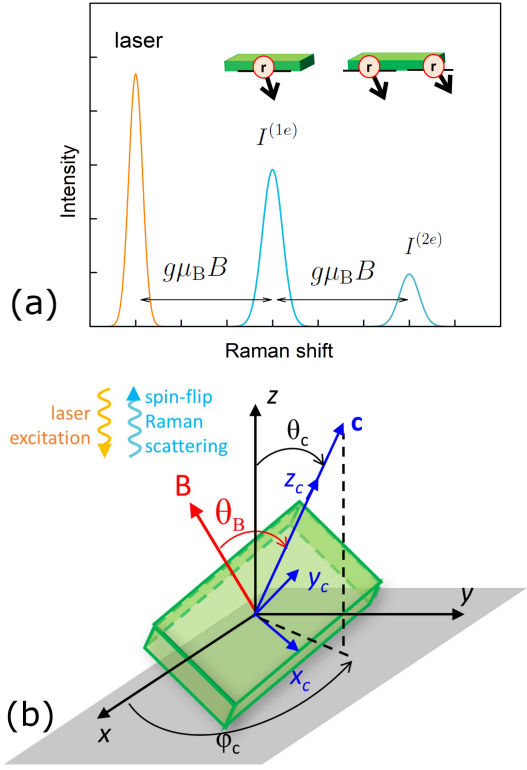}
 	\caption{(a)  Schematic illustration  of the Raman spectra (Stokes processes) with single ($I^{(1e)}$) and double ($I^{(2e)}$) SFRS signals which can be observed in the NPLs with at least one and two resident electrons, respectively.   (b) Geometry of the experiment: the $x$, $y$ and $z$-axes represent the laboratory frame, ${\textbf{c}}$ is the unit vector normal to the NPL surface, the axes $x_c$, $y_c$ and $z_c \parallel {\textbf{c}}$ represent the  NPL frame,  red arrow shows the direction of the external magnetic field. The excitation beam propagates along the $z$-axis and the SFRS signal is measured in the back scattering direction. } \label{geometry}
 \end{figure}
The schematics  of the single and double SFRS spectra and the set-up of the scattering processes under consideration are reproduced in Fig.~\ref{geometry}. The $z$ axis of the laboratory frame of reference $(x,y,z)$ is fixed parallel to the substrate surface normal. Without loss of generality, we choose the normal ($z$-direction) incidence of a monochromatic polarized light wave of the frequency $\omega$ and the backscattering geometry, orange and blue wavy lines in Fig.~\ref{geometry}. The unit vector ${\bm c}$ is directed along the normal to the nanoplatelet, it is defined by the polar angle $\theta_c$ and azimuth angle $\varphi_c$. As a rule, the CdSe nanoplatelets crystallize in zinc-blende structure and have Cd terminated and acetate-passivated (001) surfaces on both sides
\cite{zincblende0,zincblende,zincblende2}. The theory takes into account the anisotropy of the electron $g$ factor and an arbitrary orientation of the NPL with respect to the $z$ direction, laying flat, standing
straight or tilted on the substrate surface. To define the orientation of a NPL we use a second set of Cartesian axes $x_c, y_c, z_c$ with $z_c \parallel {\bm c}$. Unless otherwise specified, we extend the $D_{2d}$ point symmetry of the platelet lattice to the axial symmetry in which case all orientations of the rectangular axes $x_c$ and $y_c$ in the plane of the NPL are equivalent.

The Zeeman spin splitting of the resident electron states is described by two components of the electron $g$ factors, $g_{z_c z_c} \equiv g_{\parallel}$ and $g_{x_c x_c} = g_{y_c y_c} \equiv g_{\perp}$. In the magnetic field ${\bm B}$ making the angle $\Theta_B$ with the normal ${\bm c}$
the splitting is given by $g \mu_{\rm B} B$, where $\mu_{\rm B}$ is the Bohr magneton and $g$ is the effective electron Land\'{e} factor
\begin{eqnarray} \label{effgfactor}
g &=& \sqrt{g_\perp^2 \sin \Theta_{B}^2+g_\parallel^2  \cos \Theta_{B}^2 }\\ &=& \sqrt{g_\perp^2 +(g_\parallel^2 -g_\perp^2) ( \textbf{b} \cdot \textbf{c})^2 } \, , \nonumber
\end{eqnarray}
${\bm b} = {\bm B}/B$ is the unit vector along the magnetic field. In accordance to the experimental data of Ref.~\cite{Kudlacik2020} we take the values of $g_\perp$ and $g_\parallel$ both positive. With allowance for the $g$ factor anisotropy the spin eigenstates are oriented along and against the unit vector
\begin{equation} \label{tildeb}
\mathbf{\tilde b} = \left( g_{\perp} \textbf{b}_{\perp}  + g_{\parallel} \textbf{b}_{\parallel}\right)/g
\:,
\end{equation}
where $\textbf{b}_{\perp}$ and $\textbf{b}_{\parallel}$ are the in- and off-plane components of the vector ${\bm b}$. { In the following we denote the eigenspinors by $\pmb{\downarrow}$ and $\pmb{\uparrow}$. In the spin flip scattering observed in CdSe NPLs \cite{Kudlacik2020}, the resident electron spin reverses from $\pmb{\downarrow}$ to $\pmb{\uparrow}$ (Stokes process) or from $\pmb{\uparrow}$ to $\pmb{\downarrow}$ (anti-Stokes process).

In the next section we use the expression
\begin{equation} \label{column}
\pmb{\downarrow}\hspace{1 mm}= A_{-+}\uparrow_{\bm c} + A_{--}\downarrow_{\bm c}\:, \: \pmb{\uparrow}\hspace{1 mm} = A_{++}\uparrow_{\bm c} + A_{+-}\downarrow_{\bm c}
\end{equation}
of these two-component columns via the spin-up and spin-down states $\uparrow_{\bm c}, \downarrow_{\bm c}$ attached to the ${\bm c}$ axis. Since the spin states (\ref{column}) are orthonormalized the four coefficients $A_{ij}~(i,j = \pm)$ satisfy the identities
\begin{eqnarray} \label{ortho}
&&| A_{++}|^2 + |A_{+-}|^2 = |A_{-+}|^2 + |A_{--}|^2 =1\:,  \nonumber\\
&& A_{++}^* A_{-+} = - A_{+-}^* A_{--}  \:, \:\left\vert A_{++}^* A_{-+} \right\vert = \frac{1}{2} \sin{\tilde{\Theta}}\:, \\
&&  |A_{++}|^2 -  |A_{+-}|^2 = |A_{--}|^2 -  |A_{-+}|^2 =  \cos{\tilde{\Theta}}= \tilde{\bm b} \cdot{\bm c}\:,  \nonumber
\end{eqnarray}
where $\tilde{\Theta}$ is the angle between the vectors ${\bm c}$ and $\tilde{\bm b}$.
It follows from the definition (\ref{effgfactor}) that $ \cos{\tilde{\Theta}}= (g_{\parallel}/g)\cos \Theta_{B}$ and  $ \sin{\tilde{\Theta}}= (g_\perp/g)\sin \Theta_{B}$.

In the NPLs of the 3 to 5 monolayers thickness, the size quantization along the $z$ axis is very strong allowing us to consider only the two-dimensional envelope functions for the lateral in-plane states of the resident electrons. We use here the notation $\phi({\bm \rho})$ for the scalar envelope of the resident electron at the in-plane quantum size level and   $\phi_r({\bm \rho})$ for the resident electron localized at a NPL defect (presumably, near the platelet edge). In both cases, the lowest energy state  of the resident electron in the external magnetic field is the spin-down $\pmb{\downarrow}$ state.

The observed double SFRS \cite{Kudlacik2020} can be understood only by assuming the presence
of at least two resident electrons in a NPL. The pair of electrons cannot be unlocalized and occupy the same lowest size-quantized level in the NPL because, in that case, their ground state would have been spin singlet, see e.g. Ref.~\cite{Glazov}, and the scattering with the photon energy shift by $2 g \mu_B B$ would have been impossible. Therefore, our model implies an existence of two resident electrons localized at different in-plane cites in the platelet. We use  the notations $\phi_{1}({\bm \rho}_1)$ and $\phi_{2}({\bm \rho}_2)$ for the  in-plane scalar  envelopes of the first and second localized resident electrons.  The overlap between $\phi_{1}$ and $\phi_{2}$ is assumed to be weak so that the singlet-triplet energy splitting of the resident electrons is much smaller than the Zeeman energy in moderate magnetic fields. Thus, the lowest two-electron spin state in the external magnetic field ${\bm B}$ is the triplet double-down pair $\pmb{\downarrow} \hspace{-1 mm}\mbox{}_1\hspace{-1 mm} \pmb{\downarrow} \hspace{-1 mm}\mbox{}_2\hspace{-1 mm}$~~with spins oriented along $\tilde{\bm b}$.

Both single and double spin flip scattering processes observed in CdSe NPLs \cite{Kudlacik2020} involve the photoexcitation of the excitons.  In the model under study the incident photon generates an optically-allowed (bright) exciton formed by a heavy hole with the angular momentum projection $j_{z_c} = \pm 3/2$ on the $z \parallel {\bm c}$  axis. According to the selection rules, the optically active are the bright  NPL excitonic states with the angular momentum component $\pm 1$ along the $z_c$ axis and the envelopes
\begin{equation} \label{pm}
\Psi_{+1} = \downarrow_{\bm c} \Uparrow \Phi_{\rm exc}({\bm \rho}, {\bm \rho}_h)\:,\: \Psi_{-1} = \uparrow_{\bm c} \Downarrow \Phi_{\rm exc}({\bm \rho}, {\bm \rho}_h) \, ,
\end{equation}
where the double arrows $\Uparrow$ and $\Downarrow$ represent the heavy hole states with $j_{z_c} = 3/2$ and $j_{z_c} = - 3/2$, respectively, $\Phi_{\rm exc}({\bm \rho}, {\bm \rho}_h)$ is the two-particle scalar envelope, ${\bm \rho} =(x_c, y_c) $ and ${\bm \rho}_h = (x_{h,c}, y_{h,c})$ are the electron and hole in-plane coordinates. In the absence of any resident electrons, the matrix elements of the optical transitions into the states $\Psi_{\pm 1}$  are given by
\begin{equation} \label{meexc}
M_{\pm1}^{(\rm abs)}{(X,{\bm e}_0) {\cal E}^0} = - d_{\rm cv} \frac{e^0_{x_c} \mp {\rm i} e^0_{y_c}}{\sqrt{2}} {\cal I}_{\Phi} {\cal E}^0\:,
\end{equation}
where ${\cal E}^0$ and ${\bm e}^0$ are  the amplitude and the  polarization unit vector of the initial light electric field,
\begin{equation} \label{calPHY}
{\cal I}_{\Phi} = \int \Phi_{\rm exc}({\bm \rho}, {\bm \rho}) d {\bm \rho}\:,
\end{equation}
$d_{\rm cv}$ is the interband matrix element of the dipole moment operator, $e \langle S | x | X \rangle = e \langle S | y | Y \rangle $, with $S, X, Y$ being the Bloch functions at the $\Gamma$ point of the Brillouin zone. In general, the incident light generates in each arbitrarily oriented NPL the exciton~\cite{ELbook}
\begin{eqnarray} \label{exciton}
&&\hspace{ 1.5cm}\Psi = \left(e^0_{x_c} \Psi_{x_c} + e^0_{y_c} \Psi_{y_c} \right) \:, \\ &&  \Psi_{x_c} = \frac{\Psi_{+1}+\Psi_{-1}}{\sqrt{2}}\:,\: \Psi_{y_c} = {- {\rm i}}  \frac{\Psi_{+1}-\Psi_{-1}}{\sqrt{2}} \:. \nonumber
\end{eqnarray}
The dark exciton states described by the functions
\begin{equation} \label{pm2}
\Psi_{+2} = \uparrow_{\bm c} \Uparrow \Phi_{\rm exc}({\bm \rho}, {\bm \rho}_h)\:,\: \Psi_{-2} = \downarrow_{\bm c} \Downarrow \Phi_{\rm exc}({\bm \rho}, {\bm \rho}_h) \,
\end{equation}
do not interact with light in the absence of the resident electrons or spin-flip inducing perturbations.

The bright-dark exchange splitting $\Delta E_{\rm AF}$ is assumed to be much larger than the Zeeman energies and the uncertainty of the exciton level $\hbar \Gamma$
\begin{equation} \label{BG}
g \mu_B B \ll \hbar \Gamma \ll \Delta E_{\rm AF} \:.
\end{equation}
The same condition is also valid for the Zeeman splitting of the exciton states $g_{\rm ex}\mu_B B \cos \Theta_{B}$, where $g_{\rm ex}$ is the longitudinal component of the exciton $g$-factor (the transverse component is negligible). Since a NPL has a rectangular shape with different sides, the long-range electron-hole exchange interaction results in an anisotropic splitting $\Delta_{\rm an}$ of the bright exciton sublevels \cite{AKavokin}. Mostly, in this paper we also assume this splitting to be much smaller than $\hbar \Gamma$.

The important precondition for the SFRS process is the non-vanishing overlap between the resident electron and exciton envelopes implying also the non-vanishing exchange interaction energy {$J_{ee}$} between the resident electron and electron in the exciton.  Depending on the relation between  {$J_{ee}$} and $\Delta E_{\rm AF}$, different intermediate states consisting of three or four particle  can be formed. We consider first the single SFRS for the arbitrarily  relation between  {$J_{ee}$} and $\Delta E_{\rm AF}$. Then we consider single and double SFRS  processes in the two limiting cases of strong and weak electron-electron exchange interaction.  The proposed model is as much simplified as possible but still reproduces the main features of the spin flip Raman scattering phenomenon.

\section{Single spin flip Raman scattering} \label{single}
The equation for the efficiency of Stokes single SFRS can be written in the following form
\begin{eqnarray}  \label{intensity}
I^{(1e)} \propto |V_{f,i}^{(1e)} |^2 \delta(\hbar \omega_0 - \hbar \omega - g \mu_{\rm B} B) f_{\pmb{\downarrow}} \:.
\end{eqnarray}
Here $V_{f,i}^{(1e)}$ is the compound matrix element of the scattering from the initial state $i$ to the final state $f$, $\omega_0$ and $\omega$ are the frequencies of the initial and secondary light waves, $f_{\pmb{\downarrow}}$ is the occupation of the initial spin-down state $\pmb{\downarrow}$ of the resident electron undergoing the spin flip transition,
\begin{equation} \label{occup}
f_{\pmb{\downarrow}} = \left[1 + \exp{(-g \mu_{\rm B} B/k_{\rm B} T)} \right]^{-1}\:,
\end{equation}
$k_{\rm B}$ is the Boltzmann constant and $T$ is the temperature.
For the anti-Stokes process the sign of the Zeeman term in the $\delta$-function of Eq.~(\ref{intensity}) is reversed, and the occupation  $f_{\pmb{\downarrow}}$  replaced by the occupation $f_{\pmb{\uparrow}}=1- f_{\pmb{\downarrow}}$  of the spin-up state $\pmb{\uparrow}$.

In the second-order perturbation theory the compound matrix element has the following general form
\begin{equation} \label{MEgen}
V_{f,i}^{(1e)} = {\cal E}^0\ \sum\limits_n \frac{M^{({\rm em})}_{f,n}({\bm e}) M^{({\rm abs})}_{n,i}( {\bm e}^0) }{ E_n  - \hbar \omega_0 - {\rm i} \hbar {\Gamma_n}} \:.
\end{equation}
Here $n$ is the intermediate three-particle state formed by a resident electron and a pair of photoexcited electron and hole,  $E_n$ is the excitation energy of the intermediate state. For the Stokes process, the initial state includes an incident photon of the energy $\hbar \omega_0$ with the polarization unit vector ${\bm e}^0$  and a resident electron in the spin-down state  $\pmb{\downarrow}$ with the envelope $\varphi({\bm \rho})$ {or $\varphi_r({\bm \rho})$};  the final state includes a scattered photon of the energy $\hbar \omega$ with the polarization ${\bm e}$ and the resident electron in the spin-up state $\pmb{\uparrow}$ with the same envelope. The photon absorption and emission matrix elements, $M^{ ({\rm abs}) }_{n,i}( {\bm e}^0 )$ and $M^{({\rm em})}_{f,n}({\bm e})$, depend linearly on the unit vectors ${\bm e}^0$ and ${\bm e}^*$.

\subsection{Three particle intermediate states} \label{states}
The structure of the intermediate three-particle states $n$ depends on the relative strength of the electron-electron and electron-hole exchange interaction. Neglecting the effect of the external magnetic field, one can find the energies and the wave functions of the intermediate states as the solutions of the spin Hamiltonian
\begin{widetext}
\begin{eqnarray}  \label{e-e-h}
&&{\cal \hat H}_{e\mbox{-}e\mbox{-}h} = {\cal \hat H}_{e\mbox{-}e} + {\cal \hat H}_{2e\mbox{-}h}\,  , \quad
{\cal \hat H}_{e\mbox{-}e} = - J_{ee}\ \left(\hat{\bm s}_1\cdot \hat{\bm s}_2  - \frac{1}{4} \right), \\
&&{\cal \hat H}_{2e\mbox{-}h}= - \frac{2}{3} J_{eh} a_0^2 \left[  \hat{s}_{1,z_c} \hat{j}_{z_c} \ \delta({\bm \rho}_1 - {\bm \rho}_h)   +   \hat{s}_{2,z_c} \hat{j}_{z_c} \ \delta({\bm \rho}_2 - {\bm \rho}_h) \right], \nonumber
\end{eqnarray}
\end{widetext}
where we have omitted the common energy shift of all  states, $ \hat{\bm s}_{1,2}$ are spin operators of the first and second electrons, $\hat{j}_{z_c}$ is the diagonal 2$\times$2 matrix with the components 3/2  and $-3/2$ corresponding to the projection of the heavy hole spin on the $z_c$ axis,
$J_{eh}$ is the effective energy of the electron-hole interaction, $a_0$ is the lattice constant included to have the constant ${J}_{eh}$ in energy units. In the absence of the  interaction with the resident electron,  the energy splitting,  $\Delta E_{AF}=E_{\rm A} - E_{\rm F}$, between the bright, $E_{\pm 1} \equiv E_{\rm A}$,  and dark, $E_{\pm 2} \equiv E_{\rm F}$, neutral exciton states is given by
\begin{eqnarray}
\Delta E_{AF} = J_{eh} a_0^2 \int \Phi_{\rm exc}^2({\bm \rho},{\bm \rho}) d {\bm \rho}\, .
\end{eqnarray}
In turn, the energy {$J_{ee}$} can be written as the difference
\begin{equation}  \label{Jer}
J_{ee}= E_S - E_T
\end{equation}
between the energies $E_S = J_{ee}, E_T = 0$
in the singlet ($S$) and triplet ($T$) configurations of two electron spins if the electron-hole exchange interaction is neglected. In this limit, $J_{eh} \to 0$, the eight eigenstates of the heavy-hole three-particle Hamiltonian ${\cal \hat H}_{e\mbox{-}e\mbox{-}h}$ have the form
\begin{widetext}
\begin{eqnarray} \label{PsiST}
&& \Psi_{+3/2(-3/2)}^S = \frac{1}{\sqrt{2}}\left( \uparrow_{{\bm c},1} \downarrow_{{\bm c},2} - \downarrow_{{\bm c},1} \uparrow_{{\bm c},2} \right) \Uparrow (\Downarrow) \Phi_S ({\bm \rho}_1,{\bm \rho}_2,{\bm \rho}_h)  \, ,  \\
 &&\Psi_{+3/2(-3/2)}^T = \frac{1}{\sqrt{2}}\left( \uparrow_{{\bm c},1} \downarrow_{{\bm c},2} + \downarrow_{{\bm c},1} \uparrow_{{\bm c},2} \right)\Uparrow (\Downarrow) \Phi_T ({\bm \rho}_1,{\bm \rho}_2,{\bm \rho}_h)  \, , \nonumber\\
 && \Psi_{+5/2 (-1/2)}^T =  \uparrow_{{\bm c},1} \uparrow_{{\bm c},2} \Uparrow (\Downarrow) \Phi_T({\bm \rho}_1,{\bm \rho}_2,{\bm \rho}_h) \,  ,  \nonumber \\ &&\Psi_{+1/2 (-5/2)}^T = \downarrow_{{\bm c},1} \downarrow_{{\bm c},2} \Uparrow (\Downarrow) \Phi_T({\bm \rho}_1,{\bm \rho}_2,{\bm \rho}_h) \, .    \nonumber
 \end{eqnarray}
 \end{widetext}
Here the envelopes $\Phi_S({\bm \rho}_1,{\bm \rho}_2,{\bm \rho}_h)$ and  $\Phi_T({\bm \rho}_1,{\bm \rho}_2,{\bm \rho}_h)$  are respectively symmetric and antisymmetric under the coordinate interchange ${\bm \rho}_1 \leftrightarrow {\bm \rho}_2$, they are solutions of the $e\mbox{-}e\mbox{-}h$ three-particle static Coulomb problem. The  indexes $\pm 1/2, \pm 3/2, \pm 5/2$ show the total angular momentum component $m = s_{1,z_c} + s_{2,z_c} + j_{z_c}$. The heavy-hole columns $\Uparrow$ and $\Downarrow$ in the rhs of Eqs.~(\ref{PsiST}) correspond to the states with positive and negative values of $m$.

In the general case $J_{ee} \neq 0, J_{eh} \neq 0$, the eigen energies $E_S, E_T$ are replaced by
\begin{eqnarray} \label{eigenE}
&&E_{\pm 5/2}^T= - \frac{\Delta_1}{2}\:,\: E_{\pm 1/2}^T= \frac{\Delta_1}{2}\:,\\
&&  E_{\pm 3/2}^A = \frac{1}{2}\left(J_{ee} +  \sqrt{\Delta_{eh}^2+J_{ee}^2} \right) \:, \nonumber\\ && E_{\pm 3/2}^F  = \frac{1}{2}\left(J_{ee} - \sqrt{\Delta_{eh}^2+J_{ee}^2} \right) \:, \nonumber
\end{eqnarray}
where
\begin{eqnarray}
 \Delta_{eh} = 2J_{eh} a_0^2 \iint d {\bm \rho}  d {\bm \rho}_r\Phi_{S}({\bm \rho}_r,{\bm \rho},{\bm \rho}) \Phi_T({\bm \rho}_r,{\bm \rho},{\bm \rho}) \, ,\\
 \Delta_1 = 2J_{eh} a_0^2 \iint d {\bm \rho}  d {\bm \rho}_r \Phi_T^2({\bm \rho}_r,{\bm \rho},{\bm \rho})\, . \nonumber
\end{eqnarray}
Note that in the case $J_{ee} =  0$, $\Delta_1 \approx \Delta_{eh}\approx \Delta E_{AF}$ (see Sect.~\ref{compratio}), and  the above energies of the bright and dark excitons are given  as
$E_{A} = \Delta E_{AF}/2, E_{F} = - \Delta E_{AF}/2$.

For $J_{eh} \neq 0$, four of the eigenstates,  $\Psi^T_{m}$ with $m = \pm 1/2, \pm 5/2$, retain their form while the remaining four states become linear combinations of $\Psi^S_{\pm 3/2}$ and $\Psi^T_{\pm 3/2}$ as follows
\begin{eqnarray} \label{AFPsi}
&&\Psi^A_{\pm3/2}= C^{(1)} \Psi^S_{\pm3/2} \pm C^{(2)} \Psi^T_{\pm 3/2}\:,\: \\
 &&\Psi^F_{\pm 3/2}= \mp C^{(2)} \Psi^S_{\pm 3/2} + C^{(1)} \Psi^T_{\pm 3/2}\:, \nonumber
\end{eqnarray}
where
\[
C^{(1)} = \sqrt{\frac{1 + \gamma}{2}}\:,\:C^{(2)} = \sqrt{\frac{1 - \gamma}{2}}
\]
and
\[
\gamma = \frac{J_{ee}}{ \sqrt{\Delta_{eh}^2+J_{ee}^2}}\:.
\]

\begin{figure}[h]
	\centering
	\includegraphics[width=\linewidth]{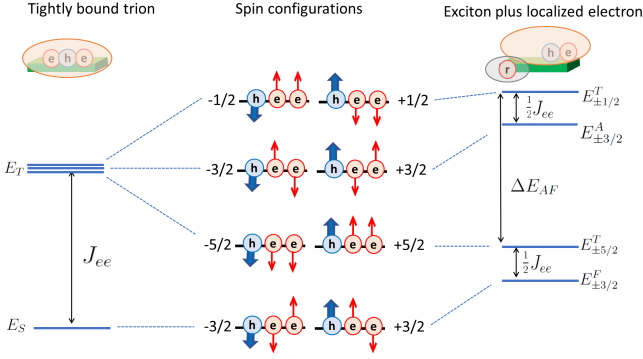}
	\caption{  Schematic illustration of the three-particle spin configurations (in the middle) and energy levels in the two limit cases: tightly bound trion ($|J_{ee}| \gg \Delta E_{AF}$, left) and ``exciton plus localized resident electron'' complex ($|J_{ee}| \ll \Delta E_{AF}$, right). The ordering of levels is shown for $J_{ee} <0$. Red thin and blue thick arrows show, respectively, the directions of the electron and hole spins relative to the ${\bm c}$ axis. 	\label{levels}  } 
\end{figure}

The states with $m = \pm 3/2$ comprise the antiparallel spins of two electrons in the singlet ($S$) or triplet ($T$) configurations mixed by the electron-hole interaction into states comprising bright ($A$) and dark ($F$)  exciton states, while the states with $m = \pm 5/2, \pm 1/2$ have the parallel triplet spin configuration of two electrons. The states with $m = \pm 5/2$   do not interact with light. The absorption and emission matrix elements for all other states are given in the Appendix \ref{AbsEmiss}. They involve  the optical overlap integrals, ${\cal I}_{S}$ and ${\cal I}_{T}$, for two electrons and a hole in the singlet and  triplet configurations, respectively}. For the estimation of  ${\cal I}_{S}$ and ${\cal I}_{T}$ one need to know the envelopes $\Phi_S$ and $\Phi_T$ of the three particle states.

While making estimations it is instructive to consider {an important special case of the complex ``exciton plus localized electron'' where the envelopes of the single resident electron and the exciton  are fixed and the Coulomb interaction between them can be treated} as a perturbation. In this case the envelopes $\Phi_{T} ({\bm \rho}_1, {\bm \rho}_2, {\bm \rho}_h) $ and  $\Phi_{S} ({\bm \rho}_1, {\bm \rho}_2, {\bm \rho}_h)$ can be presented in the ``decoupled'' form
\begin{widetext}
\begin{eqnarray} \label{psiST}
 \Phi_S({\bm \rho}_1,{\bm \rho}_2,{\bm \rho}_h)=\frac{C_S}{\sqrt{2}}\left[ \phi_r(\bm \rho_1) \Phi_{\rm exc}({\bm \rho}_2 , {\bm \rho}_h) + \phi_r(\bm \rho_2) \Phi_{\rm exc}({\bm \rho}_1 , {\bm \rho}_h) \right] \, , \\
 \Phi_T({\bm \rho}_1,{\bm \rho}_2,{\bm \rho}_h)=\frac{C_T}{\sqrt{2}} \left[ \phi_r(\bm \rho_1) \Phi_{\rm exc}({\bm \rho}_2 , {\bm \rho}_h) - \phi_r(\bm \rho_2) \Phi_{\rm exc}({\bm \rho}_1 , {\bm \rho}_h) \right] \, ,\nonumber
 \end{eqnarray}
 %\end{widetext}
where $C_{S,T}=1/\sqrt{(1\pm {\cal D}_r)}$ are additional factors differing from unity because of the overlap between the first and second functions in the brackets
%\begin{widetext}
\begin{equation} \label{overlap}
{\cal D}_r =  \iiint \phi_r({\bm \rho}_1) \phi_r({\bm \rho}_2) \Phi_{\rm exc}({\bm \rho}_1, {\bm \rho}_h)  \Phi_{\rm exc}({\bm \rho}_2, {\bm \rho}_h) d {\bm \rho}_1  d {\bm \rho}_2 d {\bm \rho}_h \, .
\end{equation}
\end{widetext}
One can show that the positive value of ${\cal D}_r$ does not exceed unity. For the functions (\ref{psiST}), the integrals (\ref{MEAF}) are reduced to
\begin{equation} \label{Itr}
{\cal I}_S = \frac{C_S}{\sqrt{2}} {({\cal I}_{\Phi} + {\cal I}_r)} \, , \quad {\cal I}_T = \frac{C_T}{\sqrt{2}} ({\cal I}_{\Phi} - {\cal I}_r)\, ,
\end{equation}
where the electron-hole overlap integral ${\cal I}_{\Phi}$ is defined by Eq.~(\ref{calPHY}) and
\begin{equation}
{\cal I}_r = \iint  \phi_r({\bm \rho}) \phi_r({\bm \rho}') \Phi_{\rm exc}({\bm \rho}, {\bm \rho}') d{\bm \rho} d {\bm \rho}'\:. \label{calI}
\end{equation}
The value of ${\cal I}_r$ as well as the values of the  overlap integrals ${\cal D}_r$ and ${\cal I}_{\Phi}$  are estimated in Section \ref{compratio}. Spin configurations of all the states are illustrated in Fig.~\ref{levels} (middle).  The structure of energy levels is shown schematically for the two limiting cases, tightly bound trion (left) and  ``exciton plus localized electron'' complex (right). Here it is worth to refer to the study of the spin-flip Raman scattering in $p$-doped GaAs/GaAlAs quantum-well structures \cite{SaCa1992,Sapega1994} where two different mechanisms were identified to contribute to the bound-hole-related scattering. The first process involves three-particle complexes, $A^0LE$, which can be considered as a localized exciton neighboring a neutral acceptor and weakly affected by the acceptor, while the second process is associated with excitons bound to neutral acceptors,  $A^0$X, acting as intermediate states.

\subsection{Single SFRS compound matrix elements} \label{matrix}
The compound matrix element (\ref{MEgen}) can be presented as a sum, $ V_T^{(1e)} + V_A^{(1e)}+ V_{F}^{(1e)}$, of three terms
\begin{widetext}
\begin{equation} \label{MEgenAT}
V_{\alpha}^{(1e)} = {\cal E}^0\ \frac{\sum\limits_{\eta = \pm } \langle f, \pmb{\uparrow} \left\vert \hat{V}^{({\rm em})}({\bm e}) \right\vert  \Psi_{\eta m_{\alpha}}^\alpha\rangle \langle  \Psi_{\eta m_{\alpha}}^\alpha \left\vert \hat{V}^{({\rm abs})}({\bm e}^0) \right\vert i, \pmb{\downarrow} \rangle}{ E^{\alpha}_{m_{\alpha}}  - \hbar \omega_0 - {\rm i} \hbar \Gamma_{\alpha,m_{\alpha}}} \, ,
\end{equation}
\end{widetext}
where $\alpha = T,A,F$ and $m_T = 1/2, m_A = m_F = 3/2$. The summation over $\pm m_{\alpha}$ in the nominator is performed by using Eqs.~(\ref{6Psi}), (\ref{em6Psi}) and results in
\begin{eqnarray} 
&&\sum\limits_{m=\pm 1/2} \langle f, \pmb{\uparrow} \left\vert \hat{V}^{({\rm em})}({\bm e}) \right\vert \Psi_{m}^T\rangle \langle  \Psi_{m}^T \left\vert \hat{V}^{({\rm abs})}({\bm e}^0) \right\vert i, \pmb{\downarrow} \rangle \nonumber  \\&&  \hspace{10 mm}=2 {\rm i} d_{\rm cv}^2 \left( e^*_{x_c} e^0_{y_c} - e^*_{y_c} e^0_{x_c} \right) A^*_{++} A_{-+} {\cal I}_{T}^2\:, \nonumber \\ &&\sum\limits_{m = \pm 3/2} \langle f, \pmb{\uparrow} \left\vert \hat{V}^{({\rm em})}({\bm e}) \right\vert  \Psi_{m}^\alpha\rangle \langle  \Psi_{m}^\alpha \left\vert \hat{V}^{({\rm abs})}({\bm e}^0) \right\vert i, \pmb{\downarrow} \rangle \nonumber\\
&& \hspace{10 mm}=  - \frac{\rm i}{2}d_{\rm cv}^2  \left( e^*_{x_c} e^0_{y_c} - e^*_{y_c} e^0_{x_c} \right) A^*_{++} A_{-+} \times \nonumber \\
&& \hspace{10 mm} \left( {\cal I}_{S} \sqrt{1\pm \gamma} \pm {\cal I}_{T}\sqrt{ 1\mp \gamma}  \right)^2 \ , \label{sums}
\end{eqnarray}
where the upper and lower signs at the right hand side correspond to $\alpha = A$ and $\alpha = F$, respectively, $|A^*_{++} A_{-+}|$ is given by Eq.~(\ref{ortho}), and  the overlap integrals ${\cal I}_{S}, {\cal I}_{T}$ are defined in Eq.~(\ref{MEAF}).
The combination $e^*_{x_c} e^0_{y_c} - e^*_{y_c} e^0_{x_c}$ is the $z_c$ component of the vector product ${\bm e}^* \times {\bm e}^0$. Therefore, the efficiency of the single SFRS depends on the scattering geometry as
\begin{equation}  \label{intensity1}
I^{(1e)} \propto \sin^2{\tilde{\Theta}} \left\vert \left( {\bm e}^* \times {\bm e}^0 \right)\cdot {\bm c} \right\vert^2\:.
\end{equation}
Importantly, the rule (\ref{intensity1}) is independent of the parameter $\gamma$ and the strength  of the electron-electron exchange interaction, while the structure of the intermediate three-particle state and the values of  ${\cal I}_{S}, {\cal I}_{T}$, of course, depend on $\gamma$. The above results can be used to obtain the single SFRS cross section for any value of $\gamma$. We will discuss further two limiting cases in order to get insight into the SFRS mechanisms involved.

\subsection{Single SFRS: two limiting cases} \label{limits}

Let us consider first the limit  of {\it strong} electron-electron exchange interaction ($|\gamma| \approx 1$) exceeding the electron-hole exchange interaction. Such the case is realized, e.g., for the exciton bound to a neutral donor, $D^0X$ complex, or a tightly bound trion state.  The negatively charged trions were shown to dominate the low-temperature PL spectra of  CdSe/CdS NPLs with thick CdS shell \cite{Shornikova2018nl} and co-exist with the exciton PL in an ensemble of bare core CdSe NPLs \cite{Shornikova2018,Kudlacik2020,Shornikova2020nn,Shornikova2020nl,Raybow2020,Ayari2020}. In this case the three-particle energies (\ref{eigenE}) transfer to
\begin{eqnarray} \label{eigenEs}
&&E^A_{\pm 3/2} \to E_S = J_{ee}\:, \\ && E_{\pm 5/2}^T, E_{\pm 1/2}^{T}, E_{\pm 3/2}^F  \to E_{T} \to 0\:,\nonumber
\end{eqnarray}
as shown at the left part of Fig.~\ref{levels}, and the matrix element (\ref{MEgen})  reduces to
\begin{eqnarray} \label{MEgenST0}
&&\hspace{0.9 cm} V_{f,i}^{(1e)}(ee) = {\rm i} d_{\rm cv}^2 {\cal E}^0 \left[\left({\bm e}^*\times{\bm e}^0 \right) \cdot {\bm c} \right] A^*_{++} A_{-+} \\ && \times \left( \frac{ {\cal I}_T^2}{ E_0  - \hbar \omega_0 - {\rm i} \hbar \Gamma_{{\rm tr},T}} -  \frac{ {\cal I}_S^2}{ E_0 + J_{ee}  - \hbar \omega_0 - {\rm i} \hbar \Gamma_{{\rm tr},S}} \right), \nonumber
\end{eqnarray}
where $E_0$ is the excitation energy of the intermediate state $n$ neglecting the exchange interactions (\ref{e-e-h}). The damping rates $\Gamma_{\rm tr}$ describe both the radiative  and nonradiative recombination processes as well as the relaxation to the lower energy levels. Since the triplet level is higher in energy  its decay rate $\Gamma_{{\rm tr},T}$ is expected to exceed $\Gamma_{{\rm tr},S}$ in which case only the singlet trion contribution to the SFRS is important.

We turn now to the limit of {\it weak} electron-electron interaction, $|\gamma| \ll 1$, or strong electron-hole exchange interaction in the exciton. In this case, the energy levels are (Fig. \ref{levels}, right)
\begin{eqnarray} \label{eigenEs2}
E_{\pm 1/2}^T = - E_{\pm 5/2}^T \approx \frac{\Delta E_{AF}}{2}\:, \hspace{4cm}\\
E^A_{\pm 3/2} \approx \frac12 (\Delta E_{AF} + J_{ee}) \:,\: E_{\pm 3/2}^F \approx - \frac12 (\Delta E_{AF} - J_{ee}) \:. \nonumber
\end{eqnarray}

For the multiplicative envelopes (\ref{psiST}), the compound matrix element is
\begin{widetext}
\begin{eqnarray} \label{MEgenST}
&&\hspace{1.3 cm} V_{f,i}^{(1e)}(eh) = {\rm i} d_{\rm cv}^2 {\cal E}^0 \left[\left({\bm e}^*\times{\bm e}^0 \right) \cdot {\bm c} \right] A^*_{++} A_{-+} \\ && \times \left[ \frac{ C_T^2 ({\cal I}_{\Phi} - {\cal I}_r)^2}{E_{A}  - \hbar \omega_0 - {\rm i} \hbar \Gamma_{A}} -  \frac{ {\cal I}_A^2}{ E_{A} + (J_{ee}/2)  - \hbar \omega_0 - {\rm i} \hbar \Gamma_{A}} - \frac{ {\cal I}_F^2}{ E_{F} + (J_{ee}/2)   - \hbar \omega_0 - {\rm i} \hbar \Gamma_{F}}\right]\, , \nonumber
\end{eqnarray}
\end{widetext}
where ${\cal I}_{\Phi}$ and ${\cal I}_r$ are defined by Eqs.~(\ref{calPHY}) and (\ref{calI}) and
\begin{eqnarray} \label{IAIF}
{\cal I}_A = \frac{{\cal I}_{\Phi} }{2}(C_S+C_T)+ \frac{{\cal I}_r}{2} (C_S-C_T),\: \\
{\cal I}_F =\frac{{\cal I}_{\Phi} }{2}(C_S-C_T)+ \frac{{\cal I}_r}{2} (C_S+C_T)\:. \nonumber
\end{eqnarray}
Again, the damping rates $\Gamma_{A,F}$ describe the recombination rates and the relaxation from the bright to dark exciton state and $\Delta E_{AF} \gg \Gamma_{A} \gg \Gamma_{F}$ \cite{Shornikova2018}.

To start the analysis of Eq.~(\ref{MEgenST})  it is worth to emphasize that, for negligibly small ${\cal I}_{r}$, $C_S = C_T = 1$ and $|J_{ee}|/2  \ll \hbar \Gamma_{A}$, Eq.~(\ref{MEgenST}) can be derived in the third-order perturbation theory. Such a theory  considers the single spin-flip scattering process as a three-stage process that involves two intermediate states $n$ and $n'$ and is described by the compound matrix element
\begin{equation} \label{gem}
V_{f,i}^{(1e)}(X,X) = {\cal E}^0\ \sum\limits_{n' n} \frac{{M^{({\rm em})}_{f,n'}(X,{\bm e}) \Delta_{n', n} M^{({\rm abs})}_{n,i}(X,{\bm e}^0)} }{ ( E_{A} - \hbar \omega_0 - {\rm i} \hbar \Gamma_{A})^2}\:.
\end{equation}
Here  {$M^{({\rm abs})}_{n,i}(X,{\bm e}^0)$ and $M^{({\rm em})}_{f,n'}(X,{\bm e}) = [M^{({\rm abs})}_{n',f}(X,{\bm e})]^*$ are, respectively, the matrix elements of the exciton generation by a photon and photon emission by an exciton {neglecting any resident electrons, see Eq.~(\ref{meexc}). In the first intermediate state $n$ the spin of the resident electron $\pmb{\downarrow}$  remains unchanged but the photon is replaced by the exciton $\Psi_{m_n}$ of Eq.~(\ref{pm}) with the angular momentum component $m_n = 1$ or $m_n = - 1$. In the second intermediate state $n'$ the resident-electron spin $\pmb{\downarrow}$ reverses to $\pmb{\uparrow}$ due to the exchange interaction $\hat{\cal H}_{e\mbox{-}e}$, see Eq.~(\ref{e-e-h}). The exchange matrix element $\Delta_{n'n}$ describes the flip-stop process. In the $e$-$e$ exchange interaction a spin flip of the electron in the exciton is neglected because of the assumed large value of the bright-dark exciton splitting. The equation for $\Delta_{n', n}$ is given by
\begin{equation}  \label{flfl2}
\Delta_{n', n} = - J_{ee} A^*_{++} A_{-+}s_{z_c} \:,
\end{equation}
where $s_{z_c} = - m_n/2$.
\newline
\subsection{Discussion of the contribution due to the integral ${\cal I}_r$} \label{indirect}
The presence of the integral ${\cal I}_r$ in Eq.~(\ref{MEgenST}) is an original result of this work and it needs an additional analysis.  Neglecting ${\cal D}_r \ll 1$ and setting $C_S = C_T = 1$ in Eq.~(\ref{IAIF}) we simplify Eq.~(\ref{MEgenST}) with ${\cal I}_A^2={\cal I}_\Phi^2$ and ${\cal I}_F^2={\cal I}_r^2$  to 
\begin{widetext}
\begin{eqnarray} \label{CSCT1}
&&\hspace{1.3 cm} V_{f,i}^{(1e)}(eh) = {\rm i} d_{\rm cv}^2 {\cal E}^0 \left[\left({\bm e}^*\times{\bm e}^0 \right) \cdot {\bm c} \right] A^*_{++} A_{-+} \\ && \times \left[ \frac{  ({\cal I}_{\Phi} - {\cal I}_r)^2}{E_{A}  - \hbar \omega_0 - {\rm i} \hbar \Gamma_{A}} -  \frac{ {\cal I}_{\Phi}^2}{ E_{A} + (J_{ee}/2)  - \hbar \omega_0 - {\rm i} \hbar \Gamma_{A}} - \frac{ {\cal I}_r^2}{ E_{F} + (J_{ee}/2)   - \hbar \omega_0 - {\rm i} \hbar \Gamma_{F}}\right]\, , \nonumber
\end{eqnarray}
\end{widetext}
The contribution proportional to $ {\cal I}_{\Phi}^2$ has been analyzed in the previous subsection, Eq.~(\ref{gem}).

Here we give an interpretation of the contributions $\propto (- 2 {\cal I}_{\Phi} {\cal I}_r)$ and $\propto {\cal I}_r^2$ in Eq. (\ref{CSCT1}). For this purpose we can set $J_{ee} =0$. In this case the level $E_A$ becomes fourfold degenerate and the denominators in the first and second terms of Eq.~(\ref{CSCT1}) coincide. While deriving Eq.~(\ref{CSCT1}) we took $C_S=C_T=1$ which allows us to take the basis of fours states at the level $E_A$ in the separable (factorized) form
\begin{eqnarray} \label{sepfact}
 &&\Psi_{A,1} = \downarrow_{{\bm c},1}\phi_r({\bm \rho}_1)  \downarrow_{{\bm c},2} \Uparrow \Phi_{\rm exc}({\bm \rho}_2,{\bm \rho}_h) \, ,    \\
&& \Psi_{A,2} = \uparrow_{{\bm c},1}\phi_r({\bm \rho}_1)  \uparrow_{{\bm c},2} \Downarrow \Phi_{\rm exc}({\bm \rho}_2,{\bm \rho}_h)  \, , \nonumber \\
 &&\Psi_{A,3} = \uparrow_{{\bm c},1}\phi_r({\bm \rho}_1)  \downarrow_{{\bm c},2} \Uparrow \Phi_{\rm exc}({\bm \rho}_2,{\bm \rho}_h) \, , \nonumber\\
 && \Psi_{A,4} = \downarrow_{{\bm c},1} \phi_r({\bm \rho}_1)  \uparrow_{{\bm c},2} \Downarrow \Phi_{\rm exc}({\bm \rho}_2,{\bm \rho}_h) \,  .  \nonumber
 \end{eqnarray}

The explanation is based on the fact that electrons are indistinguishable. Let us consider the optical generation of an exciton in the presence of the localized resident electron with the initial spin $\uparrow_{\bm c}$ or $\downarrow_{\bm c}$ oriented along ${\bm c}$. There are two channels of the generation of the states $\Psi_{A,1}$ and $\Psi_{A,2}$, with the $\sigma_+$ and $\sigma_-$ polarized light, respectively. We call them ``direct'' and ``indirect'' {and schematically illustrate in Fig.~\ref{abs}.  The first, direct, channel is standard one and can be thought just as an excitation of the four exciton states (\ref{sepfact}) as a bound state of the photoelectron and photohole. The optical matrix elements are given by Eq.~(\ref{meexc}), they are independent on the envelope $\phi_r({\bm \rho})$, and the resident electron remains unchanged. The contribution $V_{f,i}^{(1e)}(X,X) \propto {\cal I}_\Phi^2$ arises from direct processes both in absorption and emission and can also be called  the  ``geminate'' process of scattering, because the hole  recombines with the ``same'' electron that participates in the photogeneration.

However, for the transition from the initial state $i =\ \downarrow_{\bm c}$ to the intermediate state $\Psi_{A,1}$ or from the initial state $i =\ \uparrow_{\bm c}$ to the intermediate state $\Psi_{A,2}$, there is another channel in which the photohole forms the exciton state with the resident electron removed from the localization site with the envelope $\phi_r({\bm \rho})$ while the photoelectron is localized on this  localization site. The absorption matrix elements for the second, indirect, channel are
\begin{eqnarray} \label{meexc2}
&&M_{A,1}^{({\rm abs})}(X_{\rm ind}, {\bm e}_0) = \langle \Psi_{A,1} | \hat{V}^{({\rm abs})} | i = \uparrow_{\bm c} \rangle= d_{\rm cv} e^0_{\sigma_-}   \nonumber  \\ && \times \int  d{\bm \rho}\ \phi_r({\bm \rho}) \int d {\bm \rho}'  \phi_r({\bm \rho}') \Phi_{\rm exc}({\bm \rho}, {\bm \rho}')  \equiv  d_{\rm cv}e^0_{\sigma_-} {\cal I}_r \:, \nonumber\\ &&M_{A,2}^{({\rm abs})} (X_{\rm ind}, {\bm e}_0) = \langle \Psi_{A,2} | \hat{V}^{({\rm abs})} | i = \downarrow_{\bm c} \rangle = d_{\rm cv} e^0_{\sigma_+}  {\cal I}_r \:, \nonumber\\ &&
\end{eqnarray}
and similar equations are written for the emission matrix elements $M_{A,1}^{({\rm em})}(X_{\rm ind}, {\bm e}), M_{A,2}^{({\rm em})}(X_{\rm ind}, {\bm e})$.

 \begin{figure}[h]
	\centering
	\includegraphics[width=5 cm]{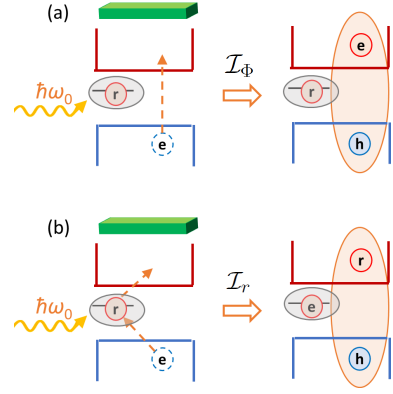}
	\caption{Illustration of (a) ``direct", $\propto {\cal I}_\Phi$, and (b) ``indirect", $\propto {\cal I}_r$,  channels of generation of the complex ``exciton plus localized resident electron''.	\label{abs} }
\end{figure}

It follows then that the contribution $\propto - 2 {\cal I}_{\Phi} {\cal I}_r$ to the SFRS arises both from the  ``direct-indirect'', 
$$
V_{f,i}^{(1e)}(X_{\rm ind},X) \propto M^{({\rm em})}_m(X_{\rm ind}, {\bm e}) M^{({\rm abs})}(X, {\bm e}_0)\:,
$$ 
and ``indirect-direct'', 
$$
V_{f,i}^{(1e)}(X,X_{\rm ind}) \propto M^{({\rm em})}_m(X, {\bm e}) M^{({\rm abs})}(X_{\rm ind}, {\bm e}_0)\:, 
$$ 
processes of the absorption and emission. The two contributions are equal providing the factor 2. For the real initial and final spin states $\pmb{\downarrow}$ and $\pmb{\uparrow}$ given by Eq.~(\ref{column}), the compound matrix elements are obtained as linear combinations of $V_{f,i}^{(1e)}(X_{\rm ind},X)$ and $V_{f,i}^{(1e)}(X,X_{\rm ind})$ which results in the factor $A^*_{++} A_{-+}$ in Eq.~(\ref{CSCT1}). Similarly, the term ${\cal I}_r^2$ comes from the indirect absorption and indirect emission. Note that this particular ``indirect-indirect''  process provides the SFRS from the dark exciton level $m_n=\pm 2$, see the last term in Eq.~(\ref{CSCT1}).

In contrast, there is no simple physical interpretation of corrections arising in Eq.~(\ref{MEgenST}), with respect to Eq.~(\ref{CSCT1}), from  the difference of the normalization factors $C_S, C_T$ from unity.  It is just worth to mention that the value of  ${\cal D}_r$ in Eq.~(\ref{overlap}) is of the same order as ${\cal I}_r^2$, see Sect. \ref{compratio}.

\section{Double spin flip Raman scattering} \label{double}
The double SFRS scattering cross section for the Stokes  process is proportional to
\begin{eqnarray}  \label{intensity20}
I^{(2e)} \propto |V_{f,i}^{(2e)} |^2 \delta(\hbar \omega_0 - \hbar \omega - 2 g \mu_{\rm B} B) f_{1,\pmb{\downarrow}} f_{2,\pmb{\downarrow}} \:.
\end{eqnarray}
Taking the same $g$-factors for the two resident electrons $j=1,2$ we can rewrite the product of occupations $f_{1,\pmb{\downarrow}} f_{2,\pmb{\downarrow}}$ as the squared occupation $f_{\pmb{\downarrow}}$ defined by Eq.~(\ref{occup}). {The both resident electron spins can flip taking into account the exchange interaction of each of them with the electron spin ${\bm s}$ in the exciton
\begin{equation} \label{e-e-e}
\hat{\cal H}_{e \mbox{-}e \mbox{-}e} = \sum_{j=1,2} J_{e_je} \left( \hat{\bm s}_j \cdot \hat{\bm s} - \frac14 \right)\:.
\end{equation}
As compared to the single SFRS, the matrix element $V_{f,i}^{(2e)}$ involves additional intermediate states and requires more complex consideration. Therefore, it is instructive to start with the calculation of $V_{f,i}^{(2e)}$ for the simplified conditions which have allowed us to derive Eq.~(\ref{gem}) and terms $\propto -2{\cal I}_\Phi {\cal I}_r$ in Eq.~(\ref{CSCT1})} for $V_{f,i}^{(1e)}$.

\subsection{Double SFRS, direct absorption and emission} \label{Doublegeminate}
{Extending the scattering mechanism involving both the direct absorption and direct emission of photons to the double SFRS in the platelets containing two resident electrons we obtain in the fourth-order perturbation approach
	\begin{widetext}
\begin{eqnarray} \label{MEdouble}
V_{f,i}^{(2e)}(X,X) &=& {\cal E}^0\ \sum\limits_{j} \sum\limits_{n'' n' n} \frac{ M^{({\rm em})}_{f,n''}(X,{\bm e}) \Delta^{(\tilde{j})}_{n'', n'} \Delta^{(j)}_{n', n} M^{({\rm abs})}_{n,i}(X,{\bm e}^0) }{ ( E_A - \hbar \omega_0 - {\rm i} \hbar \Gamma_A) ^3} \nonumber \\ &=&
\frac{ d^2_{cv}{\cal E}^0 }{E_A - \hbar \omega_0 - {\rm i} \hbar \Gamma_A}{}\left( e^*_{x_c}e^0_{x_c} + e^*_{y_c} e^0_{y_c} \right)Q {\cal I}_\Phi^2 \:.
\end{eqnarray}
\end{widetext}
Here $j = 1,2$, the index $\bar{j}$ indicates the resident electron different from the $j$-th one,
{the matrix element of the exchange interaction between the $j$-th resident electron and the electron in the exciton is defined similarly to Eq.~(\ref{flfl2}) by
\begin{equation}  \label{Dj}
\Delta^{(j)}_{n', n} = \epsilon_j s_{z_c} \:,\: \epsilon_j = - J_{e_j e} A^*_{++} A_{-+} \:,
\end{equation}
and
\begin{equation}  \label{Q}
Q = \frac12 \frac{\epsilon_1 \epsilon_2}{\left( E_{A} - \hbar \omega_0 - {\rm i} \hbar \Gamma_A \right)^2}\:.
\end{equation}}
In this case, three intermediate states $n, n'$ and $n''$ involve the same photocreated exciton and differ by the spin configuration of two resident electrons, namely, two spins down $\pmb{\downarrow}\hspace{-1 mm}_1\hspace{-1 mm} \pmb{\downarrow}\hspace{-1 mm}_2$, antiparallel spins $\pmb{\downarrow}\hspace{-1 mm}_{\bar{j}}\hspace{-1 mm} \pmb{\uparrow}\hspace{-1 mm}_j$ and two spins up $\pmb{\uparrow}\hspace{-1 mm}_1\hspace{-1 mm} \pmb{\uparrow}\hspace{-1 mm}_2$. }

The direct mechanism of the double SFRS is similar to multiple spin-flip Raman scattering observed in the diluted magnetic semiconductors, e.g. \cite{KKavokin,Furdyna,Kusrayev,Geurts,SmirnKavok}, and recently also in colloidal CdSe/CdMnS nanoplatelets \cite{Shornikova2020acs}, where ${\cal I}_r$ is certainly negligible as compared to ${\cal I}_{\Phi}$ because the $d$-electron shells are tightly bound to the Mn ions. It is also worth to mention that Eqs.~(\ref{MEdouble}), (\ref{Q}) agree with the theory of the resonance shift quantum spin noise spectroscopy and multiply spin-flip Raman scattering \cite{SmirnKavok} according to which
\[
|Q|^2 f_{\pmb{\downarrow}}^2= \frac{\left\vert \epsilon_1 \epsilon_2 \right\vert^2}{ \left\vert E_{\rm exc} - \hbar \omega_0 - {\rm i} \hbar
	\Gamma_A \right\vert^4} \langle \hat{S}_z^2(0) \hat{S}_z^2(\tau) \rangle_{2\Omega}\:,
\]
where $\hat{S}_z = \hat{s}_{1z} +  \hat{s}_{1z}$, $\hat{S}_z(\tau)$ is the Heisenberg operator,
the final factor in the right-hand side  is the Fourier transform of the correlation function $ \langle \hat{S}_z^2(0) \hat{S}_z^2(\tau) \rangle$ and equals $f_{\pmb{\downarrow}}^2/4$.

The expression in the brackets of Eq.~(\ref{MEdouble}) can be rewritten as ${\bm e}^* \cdot {\bm e}^0 - ({\bm e}^* \cdot {\bm c}) ({\bm e}^0 \cdot {\bm c})$ or, equivalently, as a scalar product of the vector products $ (\textbf{e}^*\times \textbf{c}) \cdot (\textbf{e}^0 \times \textbf{c})$, or as an explicit function of the angles $\theta_c$ and $\varphi_c$
\begin{eqnarray} \label{ecrossc}
&& \frac12 \left( 1\ +\ \cos^2 \theta_c \right) \left( e^*_{x}e^0_{x} + e^*_{y} e^0_{y} \right)  
 - \frac{ \sin^2{\theta_c}}{2} \\ && \times \left[ ( e^*_{x} e^0_{y} + e^*_{y} e^0_{x}) \sin{2\varphi_c} +(e^*_{x} e^0_{x} - e^*_{y} e^0_{y}) \cos{2\varphi_c} \right] \:. \nonumber
\end{eqnarray}
Finally, the dependence of the SFRS efficiency on the scattering geometry is given by
\begin{eqnarray}  \label{intensity2}
I^{(2e)} \propto \sin^4{\tilde{\Theta}} \left\vert {\bm e}^* \cdot {\bm e}^0 - ({\bm e}^* \cdot {\bm c}) ({\bm e}^0 \cdot {\bm c}) \right\vert^2 \:.\\
\nonumber
\end{eqnarray}

\subsection{Double SFRS, indirect absorption or emission } \label{doubelindirect}
For the mechanism related to the indirect spin-flip channel  of either absorption or emission and contributed by ${\cal I}_r$, the corresponding part of the matrix element for the Stokes process has the form
\begin{equation} \label{V2eindirect}
V_{f,i}^{(2e,{\rm ind})} = V_{f,i}^{(2e)}(X_{\rm ind},X) + V_{f,i}^{(2e)}(X, X_{\rm ind})\:,
\end{equation}
where
\begin{widetext}
\begin{equation} \label{XrX}
V_{f,i}^{(2e)} (X_{\rm ind},X) = \sum\limits_j \sum\limits_{n' n} \frac{M^{({\rm em}, \bar{j} \neq j )}_{f,n'}(X_{\rm ind},{\bm e}) \Delta^{(j)}_{n', n} M^{({\rm abs})}_{n,i}(X,{\bm e}^0) }{ ( E_A  - \hbar \omega_0 - {\rm i} \hbar \Gamma_A)^2}
\end{equation}
\end{widetext}
and the second term is obtained by the interchange $X \leftrightarrow X_{\rm ind}$.

In Eq.~(\ref{XrX}), in the first intermediate state $n$ the photon is replaced by the exciton. The exchange interaction (\ref{e-e-e}) between the electron in the exciton and  two resident electrons transforms the quantum system from the state with the initially parallel spins $\pmb{\downarrow}\hspace{-1 mm}_1 \hspace{-1 mm} \pmb{\downarrow}\hspace{-1 mm}_2$ to the second intermediate state $n'$ with a pair of the anti-parallel spins of the resident electrons and the unchanged exciton. The emission matrix element in Eq.~(\ref{XrX}) is proportional to the integral ${\cal I}_{r \bar{j}}$ given by Eq.~(\ref{calI}) in which the function $\phi_r({\bm \rho})$ is replaced by the localization envelope $\phi_{\bar{j}}({\bm \rho})$.

The summation over $m_n = \pm 1$ in the nominator of (\ref{XrX}) leads to
\begin{eqnarray} \label{V2}
&& \sum\limits_j\sum\limits_{m_n=\pm 1} M^{({\rm em}, \bar{j} \neq j)}_{f, n'}({X_{\rm ind}}, {\bm e}) \Delta^{(j)}_{n', n} M^{({\rm abs})}_{n,i}(X,{\bm e}^0) \\&&= - \frac12 d_{cv}^2  ( A^*_{++} A_{-+})^2   \left( e^*_{x_c}e^0_{x_c} + e^*_{y_c} e^0_{y_c} \right)  {\cal I}_{\Phi}  \sum\limits_j J_{e_j e} {\cal I}_{r\bar{j}} \:.\nonumber
\end{eqnarray}
The polarization and geometry dependence coincides with that for the direct double SFRS, Eq.~(\ref{intensity2}).

\subsection{Double SFRS, trion mechanism} \label{doubletrion}
{Here we consider the double SFRS with a trion as an intermediate state and assume that, in the applied magnetic field,  the initial state is formed by two spin-down resident electrons, one of them in the localized state described by the localization function $\phi_r(\rho)$ and another one in the non-localized state described by $\phi(\rho)$.  The exchange interaction between the exciton electron and non-localized resident electron is expected to be much larger than the electron-hole exchange interaction as well as the energy level uncertainties. Therefore, we consider the intermediate states $n$ and $n'$  formed by the singlet trion described by the functions $\Psi_{\pm 3/2}^S$  and the spin-down (state $n$) and spin-up (state $n'$) resident electron, respectively. Because of the singlet configuration of the electrons in the trion, the spin flip of the resident electron from $n$ to $n'$ can occur only due to the  electron-hole flip-stop exchange interaction
\begin{equation}  \label{rhexchange}
H_{e\mbox{-}h} = -\frac{2}{3}{J}_{eh} a_0^2 (  \hat{s}_{r,z_c} \hat{j}_{z_c} ) \delta({\bm \rho}_r - {\bm \rho}_h) \:,
\end{equation}
where $ \hat{s}_{r,z_c}$ is the operator of the spin projection on the ${\bm c}$ axis of the localized resident electron.}

The form of the scattering matrix element is similar to that in Eq.~(\ref{gem}):
\begin{equation}  \label{triodouble}
V_{f,i}^{(2e,{\rm tr})} = {\cal E}^0\ \sum\limits_{n' n} \frac{M^{({\rm em})}_{f,n'}({\bm e}) \Delta^{(eh)}_{n', n} M^{({\rm abs})}_{n,i}({\bm e}^0) }{( E_{{\rm tr},S} - \hbar \omega_0 - {\rm i} \hbar \Gamma_{{\rm tr},S})^2}\:,
\end{equation}
where the absorption and emission matrix elements are proportional to ${\cal I}_S$ and  given in Appendix \ref{AbsEmiss} by Eqs.~(\ref{6Psi}) and (\ref{em6Psi}), respectively. The exchange matrix element is given by
\begin{eqnarray}  \label{ehexchange2}
&&\Delta^{(eh)}_{n', n} = -\frac{2}{3}\epsilon_h  A^*_{++} A_{-+} j_{z_c}\:,\:\\
&& \epsilon_h = {J}_{eh} a_0^2 \iiint
\phi_r^2({\bm \rho}_h) \Phi_S^2({\bm \rho}_1, {\bm \rho}_2, {\bm \rho}_h) d {\bm \rho}_1 d {\bm \rho}_2 d {\bm \rho}_h\:. \nonumber
\end{eqnarray}
The estimation of ${\cal I}_S$  and $\epsilon_h$ for the trion state is given in Sect. \ref{comparioT}. The exchange energy $\epsilon_h$ between the resident electron and the hole is apparently smaller then the energy $J_{ee}$ entering the exchange matrix element for the exciton-mediated single or double SFRS. The influence of the polarization and the measurement set-up on the properties of the light scattering is identical to the previous two mechanisms.

\section{The model simplifications: Efficiency of the exciton and trion SFRS mechanisms} \label{compratio}

Here we are going to  introduce some model simplifications in order to estimate the  overlap integrals entering the trion and exciton optical transitions, electron-hole and electron-electron exchange energies, as well as to  give an estimation of the ratio between the direct to indirect matrix elements for the single SFRS.

\subsection{SFSR mediated by ``exciton plus localized electron'' complex}

We start with the representation of the three particle state as {the ``exciton} plus localized electron {complex}'' with the envelopes $\Psi_{S,T}$ described by {Eqs.}~(\ref{psiST}). If the 2D exciton Bohr radius $a^{2D}_B$ is small as compared with the in-plane dimension of the platelet then the exciton envelope function can be factorized as follows
\begin{equation}  \label{loc}
\Phi_{\rm exc}({\bm \rho}_e, {\bm \rho}_h) = f({\bm \rho}_e - {\bm \rho}_h) F\left( {\bm R}_{eh} \right) \:,
\end{equation}
where the 2D radius-vector of the exciton center of mass is given by
\begin{equation} \label{Reh}
{\bm R}_{eh} =  \frac{m_e {\bm \rho}_e + m_h {\bm \rho}_h}{m_e + m_h}\:,
\end{equation}
$m_e$ and $m_h$ are the electron and hole effective masses. The functions of translational and relative motion are approximated by
\begin{equation} \label{translate}
F\left( {\bm R}_{eh} \right) = \frac{2}{\sqrt{L_1L_2}}\cos{\frac{\pi X_{eh}}{L_1}}\cos{\frac{\pi Y_{eh}}{L_2}}\:,
\end{equation}
\begin{equation} \label{relative}
f({\bm \rho}_e - {\bm \rho}_h) = \sqrt{\frac{2}{\pi}} \ \frac{1}{\tilde{a}}\ {\rm e}^{- |{\bm \rho}_e - {\bm \rho}_h|/\tilde{a}}\:,
\end{equation}
where $L_1$ and $L_2$ are the lengths of the NPL rectangle sides along the $x_c$ and $y_c$ axes, and $X_{eh}, Y_{eh}$ are the components of ${\bm R}_{eh}$.
We assume the area $\pi \tilde{a}^2$ to be much smaller than the in-plane area of a rectangular platelet $S=L_1L_2$.  For the sake of simplicity, we take the envelope of the resident electron localized around the point ${\bm \rho}^0_r$ in the form of exponential function as well,
\begin{equation} \label{resident}
\phi({\bm \rho}_r) = \sqrt{\frac{2}{\pi}} \ \frac{1}{a^*}\ {\rm e}^{- |{\bm \rho}_r - {\bm \rho}_r^0|/a^*}\:,
\end{equation}
also assume $\pi a^{*2} \ll S$ and consider two limit cases, $\tilde{a} \ll a^*$ and  $a^* \ll \tilde{a}$.

For the assumptions (\ref{loc})--(\ref{resident}),   the overlap integral ${\cal D}_r$ in Eq. (\ref{overlap}) is approximated by
\begin{eqnarray} \label{overlap1} 
{\cal D}_r &=&  \int d {\bm \rho}_h \left[ \int d {\bm \rho}_e  \phi_r({\bm \rho}_e )  \Phi_{\rm exc}({\bm \rho}_e , {\bm \rho}_h)  \right]^2 \\
&=& 8 \pi  F^2({\bm  \rho}^0_r)\ {{\rm min} \{a^{*2}, \tilde{a}^2\}} \,, \nonumber
\end{eqnarray}
where
\begin{eqnarray} \label{overlap2}
 F^2({\bm  \rho}^0_r) = \frac{4}{S} \xi^2({\bm  \rho}^0_r)~\:,\:~\xi({\bm  \rho}^0_r)=
\cos{ \frac{\pi x^0_r}{L_1} } \cos{\frac{\pi y^0_r}{L_2}}\:.
\end{eqnarray}
While deriving Eq.~(\ref{overlap2}), and in the following estimations, we use the integral relation  (\ref{drho}) given in the Appendix \ref{B}. The optical overlap integrals, ${\cal I}_{\Phi}$ and ${\cal I}_r $, reduce to 
\begin{eqnarray} %\label{ints}
{\cal I}_{\Phi} &=& \int \Phi_{\rm exc}({\bm \rho}, {\bm \rho}) d{\bm \rho} = f(0) \int F({\bm R}) d{\bm R} = \frac{8}{\pi^2} \sqrt{ \frac{2 S}{\pi \tilde{a}^2} }\:, \nonumber \\
{\cal I}_r &=& \iint d{\bm \rho}_e d {\bm \rho}_h \phi_r({\bm \rho}_e) \phi_r({\bm \rho}_h) \Phi_{\rm exc}({\bm \rho}_e, {\bm \rho}_h) \nonumber\\
&=&2 \sqrt{2 \pi}  F({\bm  \rho}^0_r) \left\{ \begin{array}{c} ~~~\tilde{a}, ~~~~~~{\rm if}~ \tilde{a} \ll a^*\:, \\ 4 a^{*2} / \tilde{a},~~{\rm if}~ a^* \ll \tilde{a}\:. \end{array} \right. \nonumber 
\end{eqnarray}
Note  that ${\cal I}_r^2={\cal D}_r$ if $\tilde{a} \ll a^*$, and {${\cal I}_r^2 ={\cal D}_r (4a^*/\tilde a)^2$ if $a^* \ll \tilde{a}$}.
The ratio ${\cal I}_r/{\cal I}_\Phi$ is smaller than unity and can be estimated as
\begin{equation} \label{ratioI}
 \frac{{\cal I}_r}{{\cal I}_\Phi}
\sim \frac{\pi ^3 \xi({\bm \rho}^0_r)}{2S} \left\{ \begin{array}{c} \tilde{a}^2, ~~~{\rm if}~~~ \tilde{a} \ll a^*\:, \\4 a^{*2},~{\rm if}~~~ a^* \ll \tilde{a}\:. \end{array} \right.
\end{equation}

For the electron-hole exchange interaction in the three-particle complex we obtain 
\begin{eqnarray}
&&\Delta_{eh} = 2J_{eh} a_0^2 \iint \Phi_{S}({\bm \rho}_r,{\bm \rho},{\bm \rho}) \Phi_T({\bm \rho}_r,{\bm \rho},{\bm \rho}) d {\bm \rho}  d {\bm \rho}_r \nonumber\\
&&= \frac{J_{eh} a_0^2 f^2(0)}{\sqrt{1-{\cal D}_r}}\left( 1- \frac{F^2({\bm \rho}_r^0)}{f^2(0)} \right) \approx \Delta E_{AF} \, , \nonumber
\end{eqnarray}
so that in the case of weak electron-electron interaction $\gamma \approx J_{ee}/\Delta_{eh} \approx J_{ee}/\Delta E_{AF}$. In a similar way, we obtain $\Delta_1 \approx \Delta E_{AF}$.

The exchange interaction energy $J_{ee}$ entering the  spin-spin operator 
${\cal \hat H}_{e\mbox{-}e}$ can be found from
\begin{equation}  \label{Jer0}
J_{ee}=  \langle \Phi_S \vert \hat H_{\rm r,exc} \vert \Phi_S \rangle - \langle \Phi_T \vert \hat H_{\rm r,exc} \vert \Phi_T \rangle \,.
\end{equation}
The three-particle Hamiltonian reads
\begin{eqnarray}
\hat H_{\rm r,exc}= \hat T_r+\hat T_e+ \hat T_h -V_r({\bm \rho_r}) - V_r({\bm \rho_e})+V_h({\bm \rho_h}) \\
 \hspace{0.5 cm} - \,  U({\bm \rho}_r - {\bm \rho}_h) - U({\bm \rho}_r - {\bm \rho}_h) + U({\bm \rho}_r - {\bm \rho}_e) \, , \nonumber
\end{eqnarray}
where operators $\hat T$ stand for the kinetic energy of the particles, $-V_r$ and $V_h$ for the interaction of electrons and a hole with the center localizing electrons, and $\pm U$ for the two-particle repulsive and attractive interactions. Below we consider two types of the interaction, namely, the planar two-particle Coulomb potential
\begin{equation} \label{Coulomb_pot}
U_C({\bm \rho}) = \frac{e^2}{\kappa \rho}\: ,
\end{equation}
where $\kappa$ is the dielectric constant assumed to be the same inside and outside the nanostructure, and the Rytova--Keldysh potential $U_{RK}$ \cite{Rytova,Keldysh}.

 In the Heitler--London approach, see e.g. Refs.~\cite{Golub1998,Loss}, we obtain
 \begin{equation} \label{QAD}
J_{ee} = E_S - E_T = \frac{{\cal Q} + {\cal A}}{1 + {\cal D}_r} - \frac{{\cal Q} - {\cal A}}{1 - {\cal D}_r} = \frac{2 ( {\cal A} - {\cal D}_r {\cal Q})}{1 - {\cal D}_r^2}\:,
 \end{equation}
where \begin{widetext}
\begin{eqnarray} \label{JerA}
&&{\cal Q}=  \iiint \phi_r^2({\bm \rho}_r)   \Phi_{\rm exc}^2({\bm \rho}_e , {\bm \rho}_h) U({\bm \rho}_r,{\bm \rho}_e,{\bm \rho}_h) d {\bm \rho}_r  d {\bm \rho}_e d {\bm \rho}_h \:,\\ \nonumber
&& {\cal A} =  \iiint  \phi_r({\bm \rho}_r) \phi_r({\bm \rho}_e)  \Phi_{\rm exc}({\bm \rho}_r , {\bm \rho}_h)\Phi_{\rm exc}({\bm \rho}_e , {\bm \rho}_h) U({\bm \rho}_r,{\bm \rho}_e,{\bm \rho}_h) d {\bm \rho}_r  d {\bm \rho}_e d {\bm \rho}_h \: , \\ \nonumber
&& U({\bm \rho}_r,{\bm \rho}_e,{\bm \rho}_h) = U({\bm \rho}_r-{\bm \rho}_e) - U({\bm \rho}_r -{\bm \rho}_h) - V_r({\bm \rho}_e) + V_h( {\bm \rho}_h)\:.
 \end{eqnarray}
 %\end{widetext}
 
To make a long story short we present the final result for the case of two-particle Coulomb interaction $U_C$:
%\begin{widetext}
\begin{eqnarray} \label{finalJer}
&&J_{ee} =\frac{\pi F^2({\bm \rho}^0_r) \tilde{a}^2}{1 - {\cal D}_r^2}\left\{ \frac{e^2}{\kappa \tilde{a}}  \left(3 \pi - 16 \right)  +\ 16 \left[ \bar{V}_{h,\phi} - \bar{V}_{r,\phi} + F^2({\bm \rho}^0_r) (v_r - v_h)\right] \right\}\:, \hspace{1.8 cm}\mbox{if} ~~\tilde{a} \ll a^*\:,\\
&&J_{ee} = \frac{\pi  F^2({\bm \rho}^0_r) a^{*2} }{1 - {\cal D}_r^2}\left[ \frac{e^2}{\kappa a^*}  \left( 3 \pi - 32 \frac{a^*}{\tilde{a}} \right)  + \ 4 \left( 4 \bar{V}_{h,f} - \sqrt{\frac{2}{\pi}}\tilde{V}_r \right) + 16 F^2({\bm \rho}^0_r) (v_r - v_h) \right] \:, \mbox{if} ~~a^* \ll \tilde{a}\:. \nonumber
\end{eqnarray}
 \end{widetext}
Here we use the  notations for five integrals involving the potentials $V_r({\bm \rho})$ and $V_h({\bm \rho})$ given in the Appendix \ref{B}. Equations~(\ref{finalJer}) are obtained with the help of Eqs.~(\ref{drho}) and (\ref{C}).

Note that $ {\cal Q}_{r,e}$ and ${\cal Q}_{r,h}$ cancel each other and make no contribution to $J_{ee}$. The same is true for the electrostatic parts of the differences $ \bar{V}_{h,\phi} - \bar{V}_{r,\phi}$ and $v_r - v_h$.  Apparently, the terms in brackets of Eq.~(\ref{finalJer}) proportional to $F^2({\bm \rho}^0_r) \propto S^{-1}$ can be ignored in the qualitative estimations. We can make additional simplifications considering that, for bound states, the average kinetic and potential energies have the same order of magnitude and assuming the short-range parts of $V_r(\bm \rho)$ and $V_h(\bm \rho)$ to be comparable in strength. It follows then that, for $\tilde{a} \ll a^*$, the difference $\bar{V}_{h,\phi} - \bar{V}_{r,\phi}$ can be neglected as compared to $e^2/(\kappa \tilde{a})$, and for $\tilde{a} \ll a^*$, the integral $ \bar{V}_{h,f}$ and the energy $e^2/(\kappa \tilde{a})$ are smaller than $\tilde{V}_r $. Thus, for the sake of estimation, Eqs.~(\ref{finalJer}) are reduced to
\begin{eqnarray} \label{finalJer2}
J_{ee} &=&\pi \left( 3 \pi - 16 \right) F^2({\bm \rho}^0_r) \tilde{a} \frac{e^2}{\kappa}  \:,~~~~\hspace{0.5cm}\mbox{if} ~~\tilde{a} \ll a^*\:, \\
J_{ee} &=& \pi  F^2({\bm \rho}^0_r) a^* \left( 3\pi \frac{e^2}{\kappa } - 4 \sqrt{\frac{2}{\pi}}\tilde{V}_r a^*  \right) ,\:\mbox{if} ~~a^* \ll \tilde{a}\:. \nonumber
\end{eqnarray}
One can see that, in the case $\tilde{a} \ll a^*$, $J_{ee}$ is certainly negative while in the opposite case $a^* \ll  \tilde{a}$ it can change sign and  vanish at $3 \pi \sqrt{\pi} (e^2/\kappa) = 4 \sqrt{2} \tilde{V}_r a^*$.

For a thin nanostructure like a CdSe NPL, one has to take into account the difference between the large  dielectric constant $\kappa_{\rm in}$ inside the NPL and the smaller dielectric constant  $\kappa$ of the surrounding organic ligands \cite{Benchamekh2014,Shornikova2018,Ayari2020}. In this case it is preferable, instead of the potential (\ref{Coulomb_pot}), to take a more general Rytova--Keldysh potential \cite{Rytova,Keldysh} with the Fourier image 
\begin{equation} \label{V_q2}
\pm U_{RK}(q)= \pm \frac{2\pi e^2}{ \kappa q(1+qr_0)}\:,
\end{equation}
where the sign $\pm$ corresponds to the repulsive and attractive interaction, respectively, and $r_0$ is called the dielectric screening length \cite{Chernikov,Ayari2020}. For this kind of the potential, the final result for the electron-electron interaction $J_{ee}$ has the structure of Eq. (\ref{finalJer}), but with  the integrals (\ref{C}) replaced by the integrals (\ref{RK1}) containing different numerical factors $s_1, s_2$ and $s_3$. 

We are now ready to compare the direct ($\propto {\cal I}^2_\Phi$) and mixed (or direct-indirect, $\propto - 2 {\cal I}_\Phi {\cal I}_r$) exciton contributions to the single SFRS. For $\tilde{a} \ll a^*$,  their ratio at $\hbar \omega_0 = E_{A}$ can be  estimated as
\begin{equation} \label{ratio}
\frac{2|V^{(1e)}_{f,i}(X_{\rm ind},X)|}{|V^{(1e)}_{f,i}(X,X)|} \sim \frac{4 \hbar \Gamma_A }{|J_{ee}| } \frac{{\cal I}_r}{{\cal I}_\Phi}  \sim  \frac{ \hbar \Gamma_A}{ \xi({\bm \rho}^0_r)}\frac{\kappa \tilde{a}}{e^2} \:.
\end{equation}
Since the Coulomb  energy $e^2/(\kappa \tilde{a})$ is expected to exceed by far the energy uncertainty $\hbar \Gamma_A$, the indirect mechanism can be important only if the localization center lies close to the NPL boundary where a value of $|\xi({\bm \rho}^0_r)| $ is small. In the   opposite case $a^* \ll  \tilde{a}$, the value of $|J_{ee}|$ can vary in a wide range and even vanish for tightly localized states at $3 \pi \sqrt{\pi} (e^2/\kappa) = 4 \sqrt{2} \tilde{V}_r a^*$ independently of the value of  $\xi({\bm \rho}^0_r)$, in which case the scattering efficiency is reduced and the indirect mechanism becomes important. Similar considerations apply to the double SFRS.

\subsection{Estimation for the trion-mediated SFRS}\label{comparioT}

In order to estimate the scattering efficiency due to the trion intermediate  states, Eqs.~(\ref{PsiST}), we use a  factorized trion wave functions
\begin{equation} \label{trionwf}
\Phi_{S (T)}({\bm \rho}_{1}, {\bm \rho}_{2}, {\bm \rho}_h) =  f_{S(T)}({\bm \rho}_{1h}, {\bm \rho}_{2h}) F\left( {\bm R}_{\rm tr} \right) \:,
\end{equation}
where $\rho_{j h} = |{\bm \rho}_j - {\bm \rho}_h|$ ($j=1,2$), ${\bm R}_{\rm tr} = (X_{\rm tr}, Y_{\rm tr})$ are the trion center-of-mass coordinates obtained from Eq.~(\ref{Reh}) by replacing $m_e$ by $2 m_e$, and the translational motion envelope $F\left( {\bm R}_{\rm tr} \right)$ is given by the function (\ref{translate}). Here we perform estimation for the singlet-trion probe envelope function chosen in the form proposed by Chandrasekhar~\cite{Sergeev,Semina}
\begin{eqnarray} \label{probe}
&&f_{S}({\bm \rho}_{1h}, {\bm \rho}_{2h}) = {\cal K}_{S} \left[  \exp{ \left(- \frac{\rho_{1h}}{ a_1} - \frac{\rho_{2h}}{a_2} \right)} + \right. \\
&&\hspace{1 cm} \left. \exp{ \left( - \frac{ \rho_{1h} }{ a_2 } - \frac{ \rho_{2h} }{a_1} \right)}  \right]  \left( 1 + \frac{ |{\bm \rho}_1 - {\bm \rho}_2|}{a_3}\right)\:, \nonumber
\end{eqnarray}
where  $a_1, a_2$ and $a_3$ are the trial parameters, ${\cal K}_{S}$ is the normalization factor which can be approximated by $\sqrt{2}(\pi a_1 a_2)^{-1}$ and, for definiteness, we take $a_1 > a_2$. Clearly, the factorization (\ref{trionwf}) implies the conditions $\pi a_1^2, \pi a_2^2 \ll S$. The trion optical overlap integral (\ref{MEAF}) is given as
\begin{eqnarray}
&&{\cal I}_S = 2 \pi {\cal K}_{S} \left[ a_1^2  \left( 1 + \frac{2a_1}{a_3} \right) + a_2^2  \left( 1 + \frac{2a_2}{a_3} \right) \right] \nonumber \\
 &&\hspace{0.7 cm} \times  \int d{\bm \rho}\ \varphi({\bm \rho}) F({\bm \rho})\:. \nonumber
\end{eqnarray}
For a resident electron quantum-confined in the platelet, the envelope $\varphi({\bm \rho})$ is similar to the function (\ref{translate}) and the integral $\int d{\bm \rho}\ \varphi({\bm \rho}) F({\bm \rho}) = 1$. If the resident electron in the initial and final state is localized according to Eq.~(\ref{resident}), the integral  in the right-hand side has the form
$$ \int d{\bm \rho}\ \varphi_r({\bm \rho}) F({\bm \rho}) = \frac{4 \sqrt{2 \pi} a^*}{\sqrt{S}} \xi({\bm \rho}^0_r)\:.
$$
Therefore, the trion mechanism is more efficient for the single SFRS in the NPLs containing non-localized resident electrons.

The energy $\epsilon_h$ entering Eqs.~(\ref{triodouble}), (\ref{ehexchange2}) for the double SFRS  trion mechanism is estimated by
\begin{equation}
\epsilon_h \approx J_{eh} a_0^2 F^2({\bm \rho}_r^0)=\Delta E_{AF} F^2({\bm \rho}_r^0)/f^2(0) \ll \Delta E_{AF} \, .
\end{equation}
Since the electron-hole exchange energy splitting $\Delta E_{AF}$ ($\sim 4$ meV in a 4 monolayer (ML) thick CdSe NPL \cite{Shornikova2018}) is much smaller than the exciton binding energy ($\sim$ 200-300 meV in a 4 ML thick CdSe NPL \cite{Benchamekh2014,Ayari2020})}, one has $\epsilon_h \ll J_{ee}$ and the trion-involved double SFRS is less probable than the double SFRS via the exciton intermediate state.

Estimations for the triplet trion can be obtained in the similar way by using a reasonable antisymmetric probe envelope introduced by Eq.~(8) in Ref.~\cite{Semina}.

It is important to note, that in the CdSe/CdSe core-shell NPLs where the PL is dominated by the trion emission, only the single SFRS was observed \cite{Shornikova2018nl}. Both single and double SFRS was observed in the bare core CdSe NPLs under the excitation at the exciton resonance energy or slightly above~\cite{Kudlacik2020}. We conclude from the above consideration that the double SFRS involves the exciton plus two resident electrons as an intermediate state. The exchange interaction between two resident electrons is negligible as the observed Raman shift for the double SFRS is exactly 2 times the shift of the single SFRS  \cite{Kudlacik2020}. Therefore, both resident electrons are localized and at least one of them is at the edge of the NPL.  The single SFRS can be observed from the singly charged and/or doubly charged NPLs, in the latter case each of the resident electrons may contribute to the signal.

\section{Polarization selection rules and their violation} \label{polar}

In this section we discuss the polarization and geometry  selection rules for single and double SFRS and compare them with the experimental observations of Ref.~\cite{Kudlacik2020}.
 First of all, it can be noted that all considered mechanisms both for single and double SFRS do not  change the spin direction of the photocreated hole and therefore conserve  the angular momentum quantum number $m_n=+1$ or $m_n=-1$ of  the intermediate exciton state or the sign of the total angular momentum component $m$ for the three-particle intermediate states (\ref{PsiST}). Therefore, all considered mechanisms predict the co-polarization selection rules for the circular polarized light in the case when all NPLs are laying flat at the surface with $\sin \theta_c=0$ for both single and double SFRS. This can be easily seen  from the obtained polarization rules given by Eqs. (\ref{intensity1})  and (\ref{intensity2}), respectively, having in mind that $e_y = \pm i e_x$ for $\sigma_\pm$ polarized light, respectively.

For the single SFRS, the  co-polarization rules for the circular polarized light  as well as the cross-polarization  rules for the linearly polarized light  remain strict also for the whole ensemble of NPLs as
\begin{equation} \label{ecrosse}
\left( {\bm e}^* \times {\bm e}^0 \right)\cdot {\bm c} = \cos \theta_c \left( e^*_{x} e^0_{y} - e^*_{y} e^0_{x} \right) \, .   \nonumber
\end{equation}
It is important to note, that the circular co-polarization rule for the SFRS with the flip of the resident electron allows one to differentiate between  this process and the SFRS with a flip of the exciton as a whole mediated by the interaction with acoustic phonon. The latter process was observed in quantum well structures with strict circular cross-polarized selection rules in the Faraday geometry ${\bm B} \parallel z$  \cite{SaCa1992}. 

Another important difference comes for the geometry selection rules. One can easily see that the dependence of the single SFRS intensity on the NPL orientation is proportional to $\sin^2 \Theta_B \cos^2{\theta_c} = \sin^2{\theta_c} \cos^2{\theta_c}$ for the Faraday geometry. Therefore, the single SFRS with the resident electron is forbidden in the Faraday geometry for the NPLs with $\sin{\theta_c} = 0$ (horizontally laying on the surface, the face-down NPLs) as well as for the NPLs with $ \cos \theta_c = 0$ (vertically standing on the surface, the edge-up NPLs).  However, the single SFRS is allowed in the Voigt geometry ${\bm B} \perp z$ with an energy shift determined by $g=g_\perp$ for the face-down NPLs and  can be observed in the Faraday geometry from the slightly tilted  face-down NPLs (with  $g \approx g_\parallel$) and the slightly tilted edge-up NPLs (with  $g \approx g_\perp$). The analysis of the $g$-factor difference from $g_\perp$ observed in Ref.~\cite{Kudlacik2020} in the ensemble of CdSe NPLs with 4 ML thickness allowed us to conclude that the main orientations of the NPLs were face-down and slightly tilted face-down. While the polarization selection rules observed in Ref.~\cite{Kudlacik2020} for the single SFRS were mainly in agreement with the scalar triple product $\left( {\bm e}^* \times {\bm e}^0 \right)\cdot {\bm c}$, a certain violation for both the linearly and circular polarized light was observed in both the Faraday and Voigt geometries.

\subsection{Violation of the polarization selection rules for the single SFRS }

As mentioned before, the NPL in-plane anisotropy may result in an anisotropic splitting of the exciton state with eigenstates described by the functions  $\Psi_{x_c}$ and $ \Psi_{y_c}$ from Eq.~(\ref{exciton}) with the energy difference $E_{x_c}-E_{y_c}=\Delta_{\rm an}$.  Note, that $\Delta_{\rm an}$ is positive in the case $L_1 < L_2$ and can be related to the anisotropy of the long-range electron-hole exchange interaction~\cite{AKavokin}. The value of $\Delta_{\rm an}$ can be also affected by the exciton localization at anisotropic islands with the area less then $S$ \cite{Hu_Goupalov2018}. Up to now, we have neglected the effect of the anisotropic splitting assuming it to be smaller than $\hbar \Gamma$.  Its account for the three particle intermediate state ``exciton plus resident electron'' can be done within the perturbation theory as follows. Let us  consider the  perturbation $\hat H_{\rm an}$ mixing the $\pm 1$ exciton states as  $\langle \Psi_{\pm 1} \vert \hat H_{\rm an}  \vert  \Psi_{ \mp 1}  \rangle =\Delta_{\rm an}/2$.  This perturbation modifies the sum over the $m_n=\pm 1$ intermediate states and results in anisotropic  corrections $\delta V_{f,i}^{(1e,{\rm an})}(eh)$ to both the direct and mixed (or direct-indirect) exciton contributions to the single SFRS compound matrix elements   given by
\begin{eqnarray} \label{ratio_an}
 \frac{\delta V_{f,i}^{(1e,{\rm an})}(eh)}{V^{(1e)}_{f,i}(eh)} \sim {\rm i}  \frac{\Delta_{\rm an}}{2\hbar \Gamma_A}  \frac{ (e^*_{x_c} e^0_{y_c} +e^*_{y_c} e^0_{x_c})}{\left( {\bm e}^* \times {\bm e}^0 \right)\cdot {\bm c}}\ .
 \end{eqnarray}
Note, that the correction to the polarization selection rule,
 \begin{eqnarray} \label{exy}
&& e^*_{x_c} e^0_{y_c} +e^*_{y_c} e^0_{x_c}= \nonumber \\
 && \hspace{0.1 cm} \cos \theta_c   \left[(e^*_{x} e^0_{y} + e^*_{y} e^0_{x}) \cos 2\varphi_c+(e^*_{y} e^0_{y} - e^*_{x} e^0_{x}) \sin 2\varphi_c \right]   ,\nonumber
 \end{eqnarray}
depends on the NPL in-plane orientation with respect to the laboratory frame $x,y$ and the intensity of the polarized light should be averaged over the azimuth angle $\varphi_c$ in the ensemble. The anisotropic correction violates both co-polarization rule for the circularly polarized light and cross-polarization rule for the linearly polarized light for the face-down NPLs. After the averaging  over all in-plane orientations we obtain
 \begin{eqnarray} \label{int_an_av}
 \frac{I^{(1e, {\rm an})}_{\sigma^- \sigma^{+}}}{I^{(1e)}_{\sigma^+ \sigma^{+}}} =2 \frac{I^{(1e, {\rm an})}_{HH}}{I^{(1e)}_{HV}} = \frac{\Delta_{\rm an}^2}{4\hbar^2 \Gamma_A^2}\:.
 \end{eqnarray}
In the experiment \cite{Kudlacik2020}, for the nonresonant excitation the ratio $I^{(1e)}_{\sigma^- \sigma^{+}}/I^{(1e)}_{\sigma^+ \sigma^{+}}$ is estimated by 0.2 in the Faraday geometry and 1/7 in the Voigt geometry. It means that $\Delta_{\rm an}$ is of the same order as $\hbar \Gamma_A$.

Additional violations of the selection rules for the linearly polarized light in the Faraday geometry may come from the account of the  Zeeman splitting of the intermediate states in the external magnetic field neglected up to now. Indeed, in the Faraday geometry a large value of $I^{(1e)}_{HH}/I^{(1e)}_{HV}=0.5$ was observed in the magnetic field 5 T even under the nonresonant excitation \cite{Kudlacik2020}. For the ``exciton plus resident electron complex'', only the the Zeeman splitting of the exciton is important. It is controlled by the exciton effective $g$-factor and might be additionally affected by the exchange interaction of the electron in the exciton with the resident electrons not involved into the single SFRS process (when there is more than one resident electron in the NPL) or with the surface spins \cite{Shornikova2020nn}. Within the perturbation theory, it is instructive to consider the Zeeman perturbation $\hat H_Z$ leading to a splitting of the $\pm 1$ excitonic states, $\langle \Psi_{\pm 1} \vert \hat H_Z  \vert  \Psi_{ \pm 1}  \rangle = \pm \Delta_{Z}/2$.  The allowance for this perturbation  does not change the circular polarization of the emitted light but modifies the sum over the $m_n=\pm 1$ intermediate states and results in Zeeman corrections $\delta V_{f,i}^{(1e,Z)}(eh)$ to the both direct and mixed exciton contributions to the single SFRS compound matrix elements   given by
\begin{eqnarray} \label{ratio_Z}
\frac{\delta V_{f,i}^{(1e,Z)}(eh)}{V^{(1e)}_{f,i}(eh)} \sim  \frac{\Delta_{Z}}{2\hbar \Gamma_A}  \frac{ (\textbf{e}^*\times \textbf{c}) \cdot (\textbf{e}^0 \times \textbf{c})}{\left( {\bm e}^* \times {\bm e}^0 \right)\cdot {\bm c}}\ .
\end{eqnarray}
Note, that for the slightly titled face-down NPLs, $$ (\textbf{e}^*\times \textbf{c}) \cdot (\textbf{e}^0 \times \textbf{c}) \approx  e^*_{x}e^0_{x} + e^*_{y} e^0_{y}.  $$ This brings us to the crude estimation
 \begin{eqnarray} \label{int_Z}
 \frac{I^{(1e, {Z})}_{HH}}{I^{(1e)}_{HV}} = \frac{\Delta_{Z}^2}{4\hbar^2 \Gamma_A^2}
\end{eqnarray}
and   $\Delta_{Z} \sim \hbar \Gamma_A$ in the magnetic field 5 T. For the singlet trion intermediate state, the Zeeman corrections are related to the Zeeman splitting of the hole states controlled by the hole $g$-factor. However, this splitting does not violate the polarization selection rules for the circular polarizations.

\section{Conclusion} \label{conclusion}

 	To summarize, we have developed a theory of single and double SFRS observed recently in 2D nanoplatelets containing resident electrons. The derived theory is valid for arbitrary orientation of the NPLs in the ensemble as well as for arbitrary direction of the magnetic field with respect to the light incidence direction. It has been shown that the compound matrix elements for the single SFRS mediated by the trion states  or by the complexes ``exciton plus localized resident electron''  can be considered as the limiting cases of the general compound matrix elements mediated by the three-particle states with arbitrary relation between the electron-electron and electron-hole exchange interaction energies.  We have obtained the compound matrix elements  for the double SFRS for  the both limiting cases and concluded from the comparison with the experimental date that the observed SFRS signals are mediated by the excitation of  ``exciton plus localized resident electrons'' complexes. An important feature of the single-electron SFRS  distinguishing it from the exciton SFRS is the co-polarization selection rule for the circularly polarized light which can be slightly violated because of the splitting of the exciton states in the rectangularly shaped  NPLs.

The analysis of experimental data on the SFRS based on the theoretical foundation allows one to access the information about the electron $g$-factor values and anisotropy in an individual NPL as well as about the  orientation of NPLs in the ensemble \cite{Shornikova2018nl,Kudlacik2020}. Here we have additionally shown that the SFRS studies can be used as a tool to get information about the state of the resident electron in the NPL, in particular, to determine if it is localized at the edge boundary or spread over the NPL area. We have found that, for the  intermediate states ``exciton plus localized resident electron'', the indirect  photoexcitation and photorecombination channels of the scattering can play an  important role. The developed theory can be extended for the NPLs containing resident holes as well as dangling bond spins at the NPL surfaces or edges.

\section*{Acknowledgments}
 This work was funded by the Russian Foundation for Basic Research (Project 19-52-12064)  and  Deutsche Forschungsgemeinschaft (DFG) in the frame of the International Collaborative Research Center TRR 160 (Project B1).

\appendix

\section{Absorption and emission matrix elements}
\label{AbsEmiss}

Here we apply the secondary quantization method \cite{Stebe,Combes0,Combes,Glazov2020} to derive the absorption and emission matrix elements for few particles states. In this approach the operators of the electron-photon interaction for the photon absorption and emission are presented as
\begin{eqnarray} 
\hat{V}^{({\rm abs})}= - d_{cv} \left( e_{\sigma_{+}}^0  b^{\dag}_{- {\bm k}, \Uparrow}  a^{\dag}_{ {\bm k}, \downarrow  _{\bm c}} + e_{\sigma_{-}}^0   b^{\dag}_{- {\bm k}, \Downarrow}  a^{\dag}_{ {\bm k}_, \uparrow_{\bm c}} \right) c_{\hbar \omega_0, {\bm e}^0},
\nonumber\\ 
\hat{V}^{({\rm em})} = - d_{cv} \left( e_{\sigma_{+}}^*    a_{ {\bm k}, \downarrow_{\bm c} } b_{- {\bm k}, \Uparrow} + e_{\sigma_{-}}^*     a_{ {\bm k}, \uparrow_{\bm c} } b_{- {\bm k}, \Downarrow} \right) c^{\dag}_{\hbar \omega, {\bm e}}. \hspace{4 mm} \nonumber
\end{eqnarray}
Here $a^{\dag}_{{\bm k}, \uparrow_{\bm c}}, b^{\dag}_{-{\bm k}, \Uparrow}, a_{{\bm k}, \uparrow_{\bm c}}, b_{-{\bm k}, \Uparrow}$ etc. are the electron and hole creation and annihilation operators, and $c^{\dag}_{\hbar \omega, {\bm e}}, c_{\hbar \omega_0, {\bm e}^0}$ are similar operators for the photons. Hereafter we  use the notation
\begin{equation}
e_{\sigma_{\pm}}^0 =  \frac{e^0_{x_c} \mp {\rm i}  e^0_{y_c}}{\sqrt{2}}\:,\:e_{\sigma_{\pm}}^* =  \frac{e^*_{x_c} \pm {\rm i}  e^*_{y_c}}{\sqrt{2}}\:.
\end{equation}

Initial and final  states for the single SFRS are
\begin{eqnarray}
| i, \pmb{\downarrow}\rangle =  C_{\bm q} a^{\dag}_{{\bm q}, \pmb{\downarrow}} c^{\dag}_{\hbar \omega_0, {\bm e}^0 }   | 0 \rangle\:, \, 
| f, \pmb{\uparrow} \rangle = C_{\bm q} a^{\dag}_{{\bm q}, \pmb{\uparrow}} c^{\dag}_{\hbar \omega, {\bm e} }   | 0 \rangle\:,
\end{eqnarray}
where $C_{\bm q}$ is the Fourier component of the resident electron envelope.

The  intermediate states (\ref{PsiST})  in the secondary quantization approach are
\begin{eqnarray}
\Psi_{+5/2 (-1/2)}^T =  \frac{1}{\sqrt{2}} \sum\limits_{ {\bm q}_1 {\bm q}_2 {\bm q}_3} C^{(T)}_{ {\bm q}_1, {\bm q}_2, {\bm q}_3 } a^{\dag}_{{\bm q}_1, {\uparrow}_{\bm c}}  a^{\dag}_{{\bm q}_2,{\uparrow}_{\bm c}} b^{\dag}_{ {\bm q}_3,\Uparrow (\Downarrow)} | 0 \rangle , \nonumber \\
\Psi_{+1/2 (-5/2)}^T =  \frac{1}{\sqrt{2}} \sum\limits_{ {\bm q}_1 {\bm q}_2 {\bm q}_3} C^{(T)}_{ {\bm q}_1, {\bm q}_2, {\bm q}_3 } a^{\dag}_{{\bm q}_1, {\downarrow}_{\bm c}}  a^{\dag}_{{\bm q}_2,{\downarrow}_{\bm c}} b^{\dag}_{ {\bm q}_3,\Uparrow (\Downarrow)} | 0 \rangle  , \nonumber \\
\Psi_{+3/2 (-3/2)}^T =  \sum\limits_{ {\bm q}_1 {\bm q}_2 {\bm q}_3} C^{(T)}_{ {\bm q}_1, {\bm q}_2, {\bm q}_3 }  a^{\dag}_{{\bm q}_1, {\uparrow}_{\bm c}}  a^{\dag}_{{\bm q}_2,{\downarrow}_{\bm c}} b^{\dag}_{ {\bm q}_3,\Uparrow (\Downarrow)} | 0 \rangle\:  , \hspace{5 mm}\nonumber \\
\Psi_{+3/2 (-3/2)}^S =  \sum\limits_{ {\bm q}_1 {\bm q}_2 {\bm q}_3} C^{(S)}_{ {\bm q}_1, {\bm q}_2, {\bm q}_3 }  a^{\dag}_{{\bm q}_1, {\uparrow}_{\bm c}}  a^{\dag}_{{\bm q}_2,{\downarrow}_{\bm c}} b^{\dag}_{ {\bm q}_3,\Uparrow (\Downarrow)} | 0 \rangle\:  . \hspace{5 mm}\nonumber
\end{eqnarray}
Here $C^{(T)}_{ {\bm q}_1, {\bm q}_2, {\bm q}_3 } $ and $C^{(S)}_{ {\bm q}_1, {\bm q}_2, {\bm q}_3 } $ are the Fourier images of the  triplet and singlet envelopes $\Phi_T({\bm \rho}_1, {\bm \rho}_2, {\bm \rho}_h)$, $\Phi_S({\bm \rho}_1, {\bm \rho}_2, {\bm \rho}_h)$ in Eqs.~(\ref{PsiST}). The factors $1/\sqrt{2}$ in the first two equations provide the normalization condition.

Taking into account the properties of the creation and annihilation operators we can calculate the exciton absorption matrix elements. It is clear that the states $|T_{+1}, \Uparrow \rangle \equiv \Psi_{+5/2 }^T$ and  $|T_{-1} ,\Downarrow \rangle \equiv \Psi_{-5/2}^T$ do not absorb or emit photons in the dipole  approximation. For the remaining 6 matrix elements we obtain
\begin{eqnarray} \label{6Psi}
\langle \Psi_{+1/2 }^T \left\vert \hat{V}^{({\rm abs})} \right\vert i \rangle  &=& \sqrt{2} d_{\rm cv} {\cal I}_T A_{--} e^0_{\sigma_+}  \:,  \\  \langle  \Psi_{-1/2 }^T \left\vert \hat{V}^{({\rm abs})} \right\vert i \rangle &=& \sqrt{2} d_{\rm cv} {\cal I}_T A_{-+}  e^0_{\sigma_-}   \:,  \nonumber \\   \langle \Psi_{+3/2 }^S \left\vert \hat{V}^{({\rm abs})} \right\vert i \rangle  &=& d_{\rm cv}  {\cal I}_S  A_{-+} e^0_{\sigma_+}  \:, \nonumber\\    \langle  \Psi_{-3/2 }^S \left\vert \hat{V}^{({\rm abs})} \right\vert i \rangle  &=& -d_{\rm cv}  {\cal I}_S A_{--} e^0_{\sigma_-}  \:,  \nonumber\\  \langle \Psi_{+3/2 }^T \left\vert \hat{V}^{({\rm abs})} \right\vert i \rangle  &=& d_{\rm cv} {\cal I}_T A_{-+} e^0_{\sigma_+}\:, \nonumber\\   \langle \Psi_{-3/2 }^T \left\vert \hat{V}^{({\rm abs})} \right\vert i \rangle &=& d_{\rm cv} {\cal I}_T A_{--}  e^0_{\sigma_-} \:, \nonumber
\end{eqnarray}
where
 \begin{eqnarray} \label{MEAF}
{\cal I}_{S} = \sum\limits_{{\bm k} {\bm q}} C^{(S)*}_{{\bm q}, {\bm k}, - {\bm k}} C_{\bm q} = \iint \Phi_{S} ({\bm \rho}, {\bm \rho}', {\bm \rho}') \phi_r({\bm \rho})   d{\bm \rho} d{\bm \rho}', \nonumber \\ {\cal I}_{T} = \sum\limits_{{\bm k} {\bm q}} C^{(T)*}_{{\bm q}, {\bm k}, - {\bm k}} C_{\bm q} = \iint \Phi_{T} ({\bm \rho}, {\bm \rho}', {\bm \rho}') \phi_r({\bm \rho})   d{\bm \rho} d{\bm \rho}' , \nonumber\\ 
\end{eqnarray}
and the coefficients $A_{--}, A_{-+}$ are introduced in Eq. (\ref{column}). In the case of the non-localized resident electron in the initial and final states, the envelope $\phi_r({\bm \rho})$ in
Eq.~(\ref{MEAF}) should be replaced by $\phi({\bm \rho})$.

For excitation of the mixed states (\ref{AFPsi}) we have
\begin{eqnarray} \label{ISIT}
 &&  \hspace{1 cm} \langle \Psi_{\pm 3/2 }^A \left\vert \hat{V}^{({\rm abs})} \right\vert i \rangle = \\ && C^{(1)} \langle \Psi_{\pm 3/2 }^S \left\vert \hat{V}^{({\rm abs})} \right\vert i \rangle  \pm C^{(2)} \langle \Psi_{\pm 3/2 }^T \left\vert \hat{V}^{({\rm abs})} \right\vert i \rangle \:, \nonumber
\end{eqnarray}
and similar equations for $\Psi_{\pm 3/2 }^F$.

For the initial state with the resident electron in the spin-up state $\pmb{\uparrow}$, the matrix elements $\langle n | \hat{V}^{({\rm abs})} |\hspace{-1 mm} \pmb{\uparrow} \rangle$ are obtained from those in Eqs. (\ref{6Psi}) by the replacement $A_{--}, A_{-+} \to A_{+-}, A_{++}$.

The matrix elements $\langle f | \hat{V}^{({\rm em})} | n \rangle $ for the transitions $|n \rangle \to |\hspace{-1 mm}\pmb{\uparrow} \rangle$ with the photon emission are obtained from $\langle \pmb{\uparrow} | \hat{V}^{({\rm em})} |n   \rangle$ and read
\begin{eqnarray} \label{em6Psi}
\langle f \left\vert \hat{V}^{({\rm em})} \right\vert \Psi_{+1/2 }^T \rangle  &=& \sqrt{2} d_{\rm cv} {\cal I}_T A^*_{+-} e^*_{\sigma_+}  \:,  \\  \langle f \left\vert \hat{V}^{({\rm em})} \right\vert  \Psi_{-1/2 }^T \rangle &=& \sqrt{2} d_{\rm cv} {\cal I}_T A^*_{++}  e^*_{\sigma_-}   \:,  \nonumber \\   \langle f \left\vert \hat{V}^{({\rm em})} \right\vert \Psi_{+3/2 }^S \rangle  &=& d_{\rm cv}  {\cal I}_S  A^*_{++} e^*_{\sigma_+}  \:, \nonumber\\    \langle  f \left\vert \hat{V}^{({\rm em})} \right\vert \Psi_{-3/2 }^S \rangle  &=& -d_{\rm cv}  {\cal I}_S A^*_{+-} e^*_{\sigma_-}  \:,  \nonumber\\  \langle f \left\vert \hat{V}^{({\rm em})} \right\vert \Psi_{+3/2 }^T \rangle  &=& d_{\rm cv} {\cal I}_T A^*_{++} e^*_{\sigma_+}\:, \nonumber\\   \langle f \left\vert \hat{V}^{({\rm em})} \right\vert  \Psi_{-3/2 }^T \rangle &=& d_{\rm cv} {\cal I}_T A^*_{+-}  e^*_{\sigma_-} \:. \nonumber
\end{eqnarray}

\section{Integral relations and notation} \label{B}
Here we give the integral relations and explain notations used for the estimations in Sect. \ref{compratio}. The first integral relation reads 
\begin{equation} \label{drho}
\sqrt{ \frac{2}{\pi} } \frac{1}{a} \int {\rm e}^{- \rho/a} d {\bm \rho} = 2 \sqrt{2 \pi} a\:.
\end{equation}
Notations for integrals involving potentials $V_r$ and $V_h$ are
\begin{eqnarray}
&&v_r - v_h = \int \left[ V_r({\bm \rho})  - V_h({\bm \rho})\right]  d {\bm \rho}\:,\\
&& \bar{V}_{h,\phi} = \int V_h({\bm \rho}) \phi^2_r({\bm \rho})    d {\bm \rho}\:,\: \bar{V}_{r,\phi} = \int V_r({\bm \rho}) \phi^2_r({\bm \rho})    d {\bm \rho}  \:, \nonumber\\
&&\bar{V}_{h,f} = \int V_h({\bm \rho})   f^2({\bm \rho}^0_r - {\bm \rho})    d {\bm \rho}\:,\: \nonumber \\&& \tilde{V}_r = \frac{1}{a^*} \int V_r({\bm \rho}) \phi_r({\bm \rho}) d {\bm \rho}\:. \nonumber
\end{eqnarray}
Now we present integral formulas for the Coulomb potential  $U_C$
\begin{eqnarray} \label{C}
&&\frac{2}{\pi a^3} \iint U_C({\bm \rho}_e - {\bm \rho}_r) {\rm e}^{- (\rho_e + \rho_r )/ a} d{\bm \rho}_e d{\bm \rho}_r =\frac{3 \pi^2}{2} \frac{e^2}{\kappa}\:,  \nonumber \\
&&\hspace{7 mm}\int  U_C({\bm \rho})  f({\bm \rho}) d{\bm \rho} = 2 \sqrt{2\pi} \frac{e^2}{\kappa} \:, \\
&&\hspace{7 mm} \int  U_C({\bm \rho})  f^2({\bm \rho}) d{\bm \rho} = \frac{2 e^2}{\kappa \tilde{a}} \:,  \nonumber 
\end{eqnarray}
and for the   Rytova--Keldysh potential  $U_{\rm RK}$
\begin{eqnarray} \label{RK1}
&&\frac{2}{\pi a^3}  \iint U_{\rm RK}({\bm \rho}_e - {\bm \rho}_r)
{\rm e}^{- (\rho_e + \rho_r )/ a} d {\bm \rho}_e  d {\bm \rho}_r = \frac{e^2}{\kappa}s_1  \:, \nonumber \\ && \int   U_{RK}({\bm \rho})  f({\bm \rho}) d{\bm \rho} = \frac{e^2}{\kappa} s_2 \:, \\
 && \int   U_{RK}({\bm \rho})  f^2({\bm \rho}) d{\bm \rho} = \frac{e^2}{\kappa \tilde{a}} s_3   \, . \nonumber
\end{eqnarray}
Here
\begin{eqnarray}
s_1 = s^0_1  \ell  \left[ -  \frac 34 - \ln{ \ell} + \frac{\pi}{16} \ell \left( 15 + 10  \ell^2 + 3  \ell^4 \right) -  \ell^2 - \frac14  \ell^4  \right] , \nonumber
\\
s^0_1 = \frac{8\pi}{1 +  \ell^2 \left[ 3 +  \ell^2 \left( 3 +  \ell^2 \right)\right]}, \hspace{1 cm}  \ell = \frac{a}{r_0} \, . \hspace{1 cm}   \nonumber
\end{eqnarray}
 For the unscreened Coulomb interaction ($r_0 \to 0$), the coefficient $s_1$ tends to $3 \pi^2/2$. The other two factors are defined as follows
\begin{eqnarray} \label{s2}
&&s_2 = \sqrt{\frac{2}{\pi}} \eta \int\limits_{\eta}^{\infty} \frac{dx}{x R^{3/2}}  = \sqrt{\frac{2}{\pi}} \eta \left[ G(\infty,\eta) - G(\eta, \eta) \right] \:,  \nonumber \\
&&s_3 =\frac{\eta}{2\pi} \left[ G(\infty,\frac{\eta}{2}) - G(\frac{\eta}{2}, \frac{\eta}{2}) \right]\:, \quad \eta = \frac{\tilde{a}}{ r_0}\:,
\end{eqnarray}
where $R = 1 + \eta^2 - 2 \eta x + x^2$ and
\begin{eqnarray}
&&G(x, \eta)  =  \frac{  \eta x + 1 - \eta^2}{(1 + \eta^2) \sqrt{R}} -  \frac{1}{\sqrt{(1 + \eta^2)^3}} \nonumber \\ && \hspace{1 cm} \times \ln{\frac{2 (1 + \eta^2 - \eta x + \sqrt{(1 + \eta^2) R})}{x}}\:. \nonumber
\end{eqnarray}
To find the definite integral in Eq.~(\ref{s2}) we have used the indefinite integral of the function $1/(x R^{3/2})$ presented, e.g., in the book \cite{Grad}.

\end{document}